\documentclass{aa}
\usepackage{natbib}
\usepackage{graphicx}
\usepackage{amsmath}
\usepackage[psdextra,pdfencoding=auto,breaklinks=false,colorlinks=true,citecolor=blue]{hyperref}
\bibpunct{(}{)}{;}{a}{}{,}

\begin{document}

\title{A significant mutual inclination between the planets within the $\pi$ Mensae system}
\author{Robert J. De Rosa\inst{1} \and Rebekah Dawson\inst{2} \and Eric L. Nielsen\inst{3,4}}
\institute{European Southern Observatory, Alonso de C\'{o}rdova 3107, Vitacura, Santiago, Chile \\ \email{rderosa@eso.org} \and
Department of Astronomy and Astrophysics, Center for Exoplanets and Habitable Worlds, The Pennsylvania State University, University Park, PA 16802, USA \and Kavli Institute for Particle Astrophysics and Cosmology, Stanford University, Stanford, CA 94305, USA \and Department of Astronomy, New Mexico State University, P.O. Box 30001, MSC 4500, Las Cruces, NM 88003, USA}

\date{26 May 2020}

\abstract{Measuring the geometry of multi-planet extrasolar systems can provide insight into their dynamical history and the processes of planetary formation. Such measurements are challenging for systems detected through indirect techniques such as radial velocity and transit, having only been measured for a handful of systems to-date.}
{We aimed to place constraints on the orbital geometry of the outer planet in the $\pi$ Mensae system, a G0V star at 18.3\,pc host to a wide-orbit super-jovian ($M\sin i = 10.02\pm0.15$\,$M_{\rm Jup}$) with a 5.7-year period and an inner transiting super-earth ($M=4.82\pm0.85$\,$M_\oplus$) with a 6.3-d period.}
{The reflex motion induced by the outer planet on the $\pi$ Mensae star causes a significant motion of the photocenter of the system on the sky plane over the course of the 5.7-year orbital period of the planet. We combined astrometric measurements from the {\it Hipparcos} and {\it Gaia} satellites with a precisely determined spectroscopic orbit in an attempt to measure this reflex motion, and in turn constrain the inclination of the orbital plane of the outer planet.}
{We measured an inclination of $i_b=49.9_{-4.5}^{+5.3}$\,deg for the orbital plane of $\pi$~Mensae~b, leading to a direct measurement of its mass of $13.01_{-0.95}^{+1.03}$\,$M_{\rm Jup}$. We found a significant mutual inclination between the orbital planes of the two planets; a 95\% credible interval for $i_{\rm mut}$ of between $34\fdg5$ and $140\fdg6$ after accounting for the unknown position angle of the orbit of $\pi$ Mensae c, strongly excluding a co-planar scenario for the two planets within this system. All orbits are stable in the present-day configuration, and secular oscillations of planet c's eccentricity are quenched by general relativistic precession. Planet c may have undergone high eccentricity tidal migration triggered by Kozai-Lidov cycles, but dynamical histories involving disk migration or {\it in situ} formation are not ruled out. Nonetheless, this system provides the first direct evidence that giant planets with large mutual inclinations have a role to play in the origins and evolution of some super-Earth systems.}
{}

\maketitle

\section{Introduction}
$\pi$ Mensae ($\pi$ Men) is a G0V star \citep{Gray:2006ca} at a distance of 18.2\,pc. The star is known to host two planetary-mass companions; a massive super-jovian with a 5.7-year period discovered via Doppler spectroscopy \citep{Jones:2002cq}, and a transiting super-earth with a 6.3-day period \citep{Gandolfi:2018cg, Huang:2018dg}, the first discovered by the {\it Transiting Exoplanet Survey Satellite} ({\it TESS}; \citealp{Ricker:2015ie}). Combined radial velocity and transit surveys \citep{Zhu:2018js,Bryan:2019em} and the identification of long-period transiting planets \citep{her19} have shown that such system configurations are common; the majority of systems with a long-period gas giant ($a>1$\,au) also harbor short-period super-earths (1--4 $R_{\oplus}$), implying potential links between their formation and dynamics. Outer gas giants -- particularly those with large eccentricities and/or mutual inclinations -- can limit the masses and orbits super-Earths form on (e.g., \citealt{wal11,chi19}), stir up their eccentricities and mutual inclinations (e.g., \citealt{han17,hua17,lai17}), drive secular chaos (e.g., \citealt{lit11}), and in extreme cases excite their eccentricities to values high enough for tidal migration (e.g., \citealt{daw18}, Section 4.4). \citet{masu20} presented statistical evidence for a population of high mutual inclination gas giants in systems with single transiting super-Earths but due to the limits of the radial velocity and transit techniques, we have yet to directly measure the mutual inclination for any individual systems.

Full orbital characterization of systems containing both outer giant planets and inner super-Earths can also aid our understanding of the origin of short period $(<10$ day) super-Earths. Like hot Jupiters (see \citealt{daw18} for a review), they may have formed {\it in situ} or arrived close to their star via disk or tidal migration. The ``Hoptune" population (planets with orbital periods less than 10 days and sizes between 2 and 6 Earth radii) in particular exhibits similarities to hot Jupiters in host star metallicity dependence, lack of other transiting planets in the same system, and occurrence rates \citep{don18}. \mbox{\citet{daw18}} argued that high eccentricity tidal migration could account for the fact that these planets did not lose their atmospheres during the star's young, active stage. Characterizing the full three-dimensional orbital architecture of individual systems containing a super-Earth accompanied by an outer gas giant can aid us in testing the high eccentricity tidal migration scenario.

The orbital inclination of the outer planet had previously been investigated using {\it Hipparcos} astrometric measurements of the host star to detect the astrometric reflex motion induced by the planet \citep{Reffert:2011ca}. This analysis yielded a plausible range for the orbital inclination of $20\fdg3$--$150\fdg6$, corresponding to a maximum mass for the companion of 29.9\,$M_{\rm Jup}$. This analysis was not able to conclusively differentiate between a co-planar and a mutually inclined configuration for the system. Efforts have also been made to resolve the outer planet via direct imaging, but the contrast achieved was not sufficient to detect the companion \citep{Zurlo:2018da}. A single epoch of relative astrometry between the planet and the host star would likely lead to an immediate determination of the inclination of the orbital plane.

In this paper we present an analysis of spectroscopic and astrometric measurements of the motion of the star induced by the orbit of $\pi$ Men b. We describe the acquisition of the data in Section~\ref{sec:acq}. We repeat the analysis presented in \citep{Reffert:2011ca} to demonstrate the improvement of the inclination constraints achieved with a revised spectroscopic orbit in Section~\ref{sec:hip-only}. We incorporate the latest astrometric measurements from {\it  Gaia} into our joint astrometric-spectroscopic model in Section~\ref{sec:gaia}. We discuss the measured mutual inclination in the context of the dynamical history of the system in Section~\ref{sec:dynamics}, and conclude in Section~\ref{sec:conclusion}.

While this manuscript was being reviewed an independent analysis of the same datasets by \citet{Xuan:2020aa} was published. We derive consistent results on the mutual inclination of the two planets despite using a different approach for incorporating the various astrometric measurements of the system.

\section{Data Acquisition}
\label{sec:acq}
\subsection{Radial velocities}
\label{sec:rv}
We collected 359 literature and archival radial velocities of the $\pi$~Men primary star were obtained between 1998 January 16 and 2020 January 8 using the University College London Echelle Spectrograph (UCLES; \citealp{Diego:1990bs}) and the High Accuracy Radial velocity Planet Searcher (HARPS; \citealp{Mayor:2003wv}). The 77 velocities from UCLES span from 1998 January 16 until 2015 November 22 (R. Wittenmyer, {\it priv. comm.}), the same record used for the discovery and characterization of $\pi$~Men~c \citep{Gandolfi:2018cg, Huang:2018dg}. The 282 velocities from HARPS span from 2003 December 28 until 2020 January 08 and were obtained from the reduced data products publicly available on the ESO Archive Facility\footnote{\url{https://archive.eso.org/}}, an additional seven velocities from HARPS were available but were rejected due to a deterioration of the thorium-argon lamp on 2018 November 26 and 27. The HARPS instrument underwent an intervention on 2015 June 03 that led to a significant change in the radial velocity zero point of the instrument \citep{LoCurto:2015vd}. To correct for this in our analysis we treat the 129 velocities from before this date as coming from a different instrument (``HARPS1'') to the remaining 153 (``HARPS2''). A complete listing of the radial velocities used in this study is given in Table~\ref{tbl:rvs}.

\subsection{Astrometric data}
\label{sec:acq-astro}
\begin{table}
\caption{Astrometric measurements of the photocenter of the $\pi$ Mensae system}
\label{tbl:astro}
 \centering
\begin{tabular}{rccl}
\hline\hline
Property & Value & Error & Unit \\
\hline
\multicolumn{4}{c}{{\it Hipparcos} (1991.25) - HIP 26394}\\
\hline
$\alpha$ & $84.28663594$ &  $0.20$ & deg, mas\tablefootmark{a,b}\\
$\delta$ & $-80.47167440$ &  $0.20$ & deg, mas\tablefootmark{a}\\
$\varpi$ & $54.60$ &  $0.21$ & mas\\
$\mu_\alpha^{\star}$ & $312.01$ & $0.24$ & mas\,yr$^{-1}$\\
$\mu_\delta$ & $1050.38$ &  $0.26$ & mas\,yr$^{-1}$ \\
\hline
\multicolumn{4}{c}{{\it Gaia} (2015.50) - Gaia DR2 4623036865373793408}\\
\hline
$\alpha$ & $84.29927749888$ &  $0.3507$ & deg, mas\tablefootmark{a,b}\\
$\delta$ & $-80.46460401033$ &  $0.1772$ & deg, mas\tablefootmark{a}\\
$\varpi$ & $54.7052$ &  $0.0754$ & mas \\
$\mu_\alpha^{\star}$ & $311.246$ & $0.148$ & mas\,yr$^{-1}$\\
$\mu_\delta$ & $1048.908$ &  $0.159$ & mas\,yr$^{-1}$\\
\hline
\end{tabular}
\tablefoot{\tablefoottext{a}{Value in degrees, uncertainty in milliarcseconds}
\tablefoottext{b}{Uncertainty in $\alpha^\star = \alpha\cos\delta$}}
\end{table}
\begin{figure}
    \centering
    \includegraphics[width=0.5\textwidth]{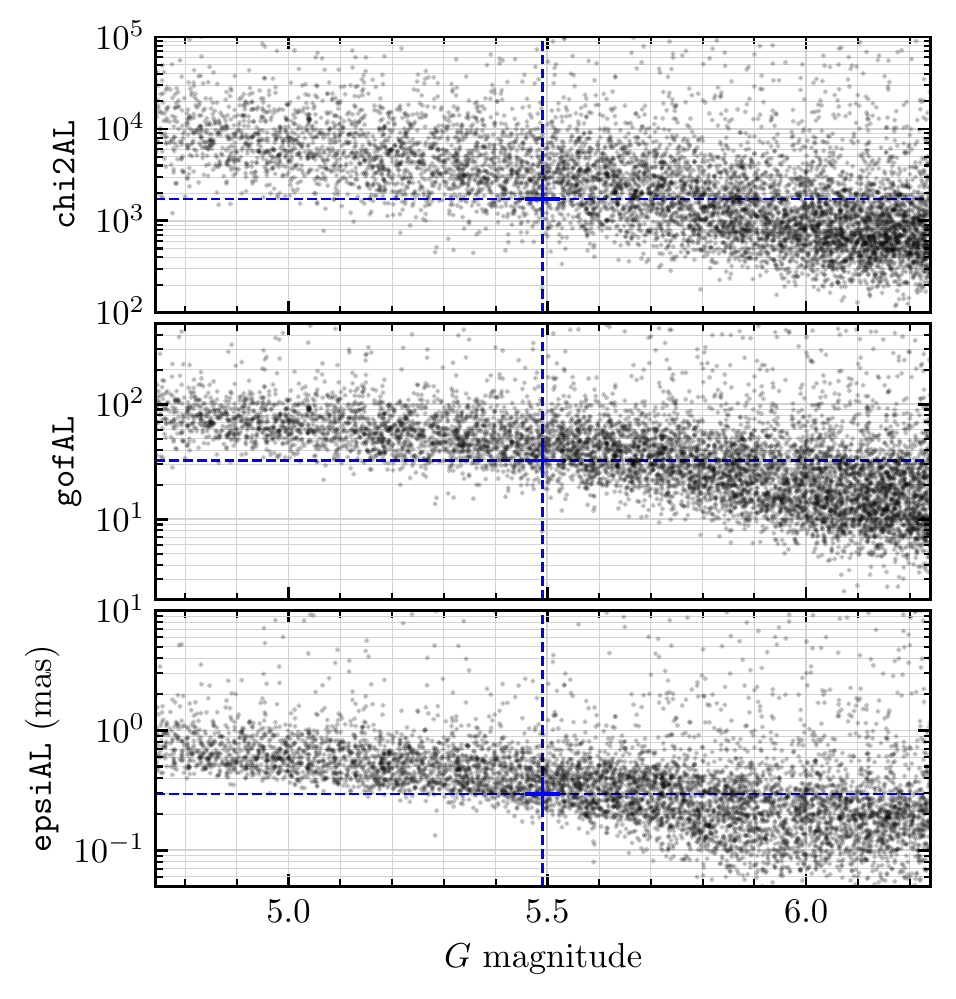}
    \caption{Quality of fit metrics for $\pi$ Men (yellow cross) and a sample of stars within a factor of two brightness (points) from the {\it Gaia} catalogue. The parameters shown are the the $\chi^2$ of the fit of the five-parameter astrometric model to the measurements (top panel), an analogue of a reduced $\chi^2$ (middle panel), and the sum of the residuals (bottom panel). The quality of the fit for $\pi$ Men is consistent with that of other stars of a similar brightness.}
    \label{fig:gaia-stats}
\end{figure}

Astrometric measurements of the $\pi$ Men system were obtained from {\it Hipparcos} catalogue re-reduction \citep{vanLeeuwen:2007dc} and from the second {\it Gaia} data release (DR2; \citealp{GaiaCollaboration:2018io}). The coordinates are expressed in the ICRS reference frame at the 1991.25 epoch for {\it Hipparcos} and the 2015.5 epoch for {\it Gaia}. A minor correction to the {\it Gaia} astrometry was applied to account for the orientation and rotation of the bright star reference (for $\pi$ Men: $\Delta\alpha^\star=-1.37\pm0.34$\,mas, $\Delta\delta=0.15\pm0.16$\,mas, $\Delta\mu_{\alpha^\star}= \Delta\mu_\delta = 0.06\pm0.05$\,mas\,yr$^{-1}$; \citealp{Lindegren:2018gy,Lindegren:2019iv})\footnote{We use the notation $\alpha^\star = \alpha\cos\delta$}, and the $\sim10\%$ error inflation terms described in \cite{Arenou:2018dp}. We use the revised orientation and spin parameters presented in \citet{Lindegren:2020ik}; $\epsilon_X=-0.019\pm0.158$\,mas, $\epsilon_Y=1.304\pm0.349$\,mas, $\epsilon_Z=0.553\pm0.135$\,mas, $\omega_X=-0.068\pm0.052$\,mas\,yr$^{-1}$, $\omega_Y=-0.051\pm0.045$\,mas\,yr$^{-1}$, $\omega_Z=-0.014\pm0.066$\,mas\,yr$^{-1}$), applying them to the catalogue values using a Monte Carlo-based approach. The goodness of fit metrics given in the {\it Gaia} catalogue are better than for a typical star of a similar magnitude (Figure~\ref{fig:gaia-stats}). The astrometric measurements from both catalogues, after applying the correction for the {\it Gaia} values, are given in Table~\ref{tbl:astro}.

In addition to the {\it Hipparcos} catalogue values, we also obtained the individual measurements made by the satellite of $\pi$ Men that were used to derive the five astrometric parameters given in Table~\ref{tbl:astro}. These measurements are effectively one-dimensional, constraining the position of the star to within 1--2\,mas along a great circle on the celestial sphere with a scan orientation angle $\psi$ (a north-south scan has $\psi=\pi/2$). Instead of publishing the abscissa data $\Lambda$---the raw scan measurements---the re-reduction of the {\it Hipparcos} data presented the abscissa residuals $\delta\Lambda$ after the best-fit astrometric model had been subtracted. These are referred to as the {\it Hipparcos} intermediate astrometric data (IAD; \citealp{vanLeeuwen:2007du}).

The IAD for $\pi$ Men contain the epoch $t$, parallax factor $\Pi$, scan orientation $\psi$, abscissa residual $\delta_\Lambda$ and corresponding uncertainty $\sigma_\Lambda$ for each of the 137 measurements of the star (Table~\ref{tbl:iad}). The measurements span from 1989 November 05 until 1993 January 29, have an average uncertainty of 1.6\,mas, and none were rejected when fitting the astrometric model presented in the catalogue. The abscissa $\Lambda_{\rm HIP}$ can be reconstructed using the formalism described in \cite{Sahlmann:2010hh} as
\begin{equation}
    \Lambda_{\rm HIP} = \left(\alpha^\star + \mu_{\alpha^\star}t\right)\cos\psi + \left(\delta + \mu_\delta t\right)\sin\psi + \varpi\Pi + \delta\Lambda
\end{equation}
We also decomposed four of the astrometric parameters into a constant term and an offset (e.g., $\alpha^\star = \alpha^\star_{\rm H} + \Delta\alpha^\star$) to avoid precision loss. When reconstructing the abscissa both the constant and offset terms were set to zero:
\begin{equation}
    \Lambda_{\rm HIP} =  \varpi\Pi + \delta\Lambda
\end{equation}
which we can compare to a noiseless model abscissa generated with a small perturbation of the catalogue parameters
\begin{equation}
\label{eq:abs}
    \Lambda = \left(\Delta\alpha^\star + \Delta\mu_{\alpha^\star}t \right)\cos\psi + \left(\Delta\delta + \Delta\mu_\delta t\right)\sin\psi + \varpi\Pi
\end{equation}
in order to compute a goodness of fit, for example.

The individual scan measurements from the {\it Gaia} satellite are only due to be published at the conclusion of the mission, limiting us to the astrometric parameters presented in the catalogue. However, the predicted date, parallax factor, and scan orientation $\theta$ of the individual {\it Gaia} measurements are available\footnote{\url{https://gaia.esac.esa.int/gost/}} (Table~\ref{tbl:gost}), giving us the necessary information to perform a simplistic simulation of the {\it Gaia} measurement of the position and motion of the photocenter. Unlike for {\it Hipparcos}, here the scan orientation describes the position angle of the scan direction relative to north\footnote{\url{https://www.cosmos.esa.int/web/gaia/scanning-law-pointings}} (i.e. a north-south scan has $\theta=0$). These simulated measurements can be compared with the catalogue values to constrain the photocenter motion during the time span of the {\it Gaia} measurements. We generated a forecast using this tool on 2020 January 7, resulting in a prediction of 26 measurements between 2014 August 26 and 2016 May 20, within the range of dates for which measurements were used to construct the DR2 catalogue. This is more than the 21 transits used to compute the astrometric parameters in the catalogue, consistent with the warning on the {\tt gost} utility website that only 80\% of the measurements are usable.

\section{Photocenter motion measured by Hipparcos}
\label{sec:hip-only}
The presence of a massive companion to $\pi$ Men makes interpretation of astrometric measurements of this object more challenging. In a single star system the barycenter and photocenter of the system are coincident with the location of the star. In an unresolved multiple system with two or more objects of unequal brightness---as is the case for the $\pi$ Men system---the photocenter, barycenter, and the location of the primary star are no longer coincident. Astrometric measurements of such a system are therefore of the position and motion of the photocenter of the system, a combination of the constant velocity of the system barycenter through space and the orbital motion of the photocenter around the barycenter.

We first investigated whether a combination of the {\it Hipparcos} IAD and the best fit spectroscopic orbit can provide any significant constraint on the geometry of the orbit of $\pi$ Men b. We implemented the grid-based approach presented in \cite{Sahlmann:2010hh} where for each combination of the orbital inclination $i$ and longitude of the ascending node $\Omega$, the orbital elements measured from the spectroscopic orbit were fixed, and the astrometric parameters were calculated using a least-squares based approach.

\subsection{Model description}
A Keplerian fit to the radial velocities provides a measurement of the period $P$, velocity semi-amplitude $K_1$, eccentricity $e$, argument of periastron $\omega_\star$ and the time of periastron $T_0$ of the orbit of the star around the system barycenter. Here we use the notation $\omega_\star$ to refer to the argument of periastron for the primary star, and $\omega = \omega_\star + \pi$ as the argument of periastron for the companion. These spectroscopic elements can be combined with an assumed mass of the primary star $M_1$ and the inclination of the orbit $i$ to determine the mass ratio $q=M_2/M_1$ by solving the following expression for the mass function
\begin{equation}
    \frac{PK_1^3}{2\pi G}\left(1-e^2\right)^{3/2} = \frac{q^3}{\left(1+q\right)^2}M_1\sin^3i
\end{equation}
where $G$ is the gravitational constant. The total semi-major axis $a$ in astronomical units can then be calculated using Kepler's third law as
\begin{equation}
    a^3 = P^2\left(M_1+M_2\right)
\end{equation}
when the period and masses are expressed in years and solar masses.

The semi-major axis of the photocenter orbit $a_p$ around the barycenter of a binary system can be calculated from the total semi-major axis of the orbit $a$, the masses of the two components $M_1$ and $M_2$, and their magnitude difference $\Delta m=-2.5\log(F_2/F_1)$. We define the fractional mass $B$ as 
\begin{equation}
B = \frac{M_2}{M_1+M_2}
\end{equation}
and the analogous fractional flux $\beta$ as
\begin{equation}
\beta = \frac{F_2}{F_1+F_2} = \left(1 + 10^{0.4\Delta m}\right)^{-1},
\end{equation}
from which $a_p$ can be calculated as $a_p = a\left(B-\beta\right)$ (e.g., \citealp{Heintz:1978wp,Coughlin:2012fz}). In a binary system where the companion is much fainter than the primary star, the value of $\beta$ asymptotes to zero and the photocenter orbit can be considered as coincident with the orbit of the primary around the barycenter (i.e. $a_p\approx a_1$). We include this term for completeness; it is only relevant for extremely low inclinations where the mass of the secondary exceeds the hydrogen burning limit. The value of $\beta$ is almost always wavelength dependent unless the spectral energy distributions of the two objects are identical. We refer to fractional fluxes computed using the {\it Hipparcos} $H_p$ filter as $\beta_H$, and those using the {\it Gaia} $G$ filter as $\beta_G$. We used an empirical mass-magnitude relationship for stars to determine the flux ratio for solutions where $M_2>0.077$\,$M_{\odot}$ \citep{Pecaut:2013ej}, and set $\beta_H=\beta_G=0$ otherwise.

The position of the photocenter relative to the barycenter in the sky plane can be calculated from the elements discussed previously, the position angle of the ascending node $\Omega$ and the epoch of the observation. The radius of a photocenter orbit $r_p$ at a given epoch is defined as
\begin{equation}
    r_p = \frac{a_p \left(1-e^2\right)}{1 + e\cos\nu}
\end{equation}
where the true anomaly $\nu$ is computed from the mean and eccentric anomalies $M$ and $E$
\begin{equation}
\begin{split}
    M &= \frac{2\pi}{P}\left(t - T_0\right) = E - e\sin E, \\
    \nu & = \cos^{-1}\left(\frac{\cos E - e}{1 - e\cos E}\right).
\end{split}
\end{equation}
The offset between the photocenter and barycenter in the right ascension $x$ and declination $y$ directions can be calculated as
\begin{equation}
\begin{split}
    x &= r_p\left[\cos\left(\omega_\star + \nu \right)\sin\Omega + \sin\left(\omega_\star + \nu \right)\cos\Omega\cos i\right] \\
    y &= r_p\left[\cos\left(\omega_\star + \nu \right)\cos\Omega - \sin\left(\omega_\star + \nu \right)\sin\Omega\cos i\right] \\
\end{split}
\end{equation}

We modeled a {\it Hipparcos} abscissa by combining the five standard astrometric parameters described previously with the motion of the photocenter in the right ascension and declination directions. As in \cite{Sahlmann:2010hh} we defined a new variable
\begin{equation}
    \Upsilon = x\cos\psi + y\sin\psi
\end{equation}
that converts the two-dimensional measurement into a one-dimensional measurement along the orientation of the scan. This variable is expressed in astronomical units, allowing us to account for a potential change in the best fit parallax of the star. We combined this new variable with our model abscissa given in Equation~\ref{eq:abs} as
\begin{equation}
\label{eq:abs_model}
    \Lambda = \left(\Delta\alpha^\star + \Delta\mu_{\alpha^\star}t \right)\cos\psi + \left(\Delta\delta + \Delta\mu_\delta t\right)\sin\psi + \varpi\left(\Pi + \Upsilon\right)
\end{equation}
The least-squares solution to this linear equation can be found numerically via a matrix-based approach:
\begin{equation}
    \begin{bmatrix}
    \Delta\alpha^\star \\
    \Delta\delta \\
    \Delta\mu_{\alpha^\star}\\
    \Delta\mu_\delta\\
    \varpi
    \end{bmatrix}
    = \left(A^T\cdot A\right)^{-1} \cdot \left(A^T\cdot\Lambda_{\rm HIP}\right),
\end{equation}
where $A$ is a (5,137) matrix constructed using the values from the {\it Hipparcos} IAD and the variable $\Upsilon$ defined previously;
\begin{equation}
A = 
    \begin{bmatrix}
    \cos \psi_0 & \sin \psi_0 & t_0\cos\psi_0 & t_0\sin\psi_0 & \Pi_0 + \Upsilon_0  \\
    \cos \psi_1 & \sin \psi_1 & t_1\cos\psi_1 & t_1\sin\psi_1 & \Pi_1 + \Upsilon_1 \\
    \vdots & \vdots & \vdots & \vdots & \vdots\\
    \cos \psi_n & \sin \psi_n & t_n\cos\psi_n & t_n\sin\psi_n & \Pi_n + \Upsilon_n \\
    \end{bmatrix}.
\end{equation}
The least-squares solution can be used to construct a model abscissa using Equation~\ref{eq:abs_model} which can be compared with the reconstructed {\it Hipparcos} abscissa to calculate the goodness of fit
\begin{equation}
    \chi^2 = \sum{\left(\frac{\Lambda - \Lambda_{\rm HIP}}{\sigma_\Lambda}\right)^2}.
\end{equation}
This calculation was performed on a finely sampled $i$-$\Omega$ grid, with 500 elements between $i=0$\,deg and $180$\,deg and 1000 elements between $\Omega=0$\,deg and $360$\,deg.

\subsection{Model validation}
We verified our model implementation by comparing to the results of a previous study combining radial velocity and {\it Hipparcos} measurements by \citet{Reffert:2011ca}. We first applied it to \object{HD 168443}, a brown dwarf with a minimum mass of $M\sin i=18$\,$M_{\rm Jup}$ on a 4.8\,yr orbit. To ensure consistency with the previous study we use the same spectroscopic orbital elements from \cite{Butler:2006dd}. We found $\chi^2_{\rm min}=65.3$ at $i=36\fdg1$ $\Omega=131\fdg1$ and a 3-$\sigma$ range for $i$ of $17\fdg8$--$163\fdg0$, consistent with their results for this system. Our $\chi^2$ surface for this example is a good match to their Figure~3. We found similarly consistent results for other systems in their study.

\subsection{Application to $\pi$ Mensae}
\label{sec:rv-fit}
\begin{figure}
    \centering
    \includegraphics[width=0.5\textwidth]{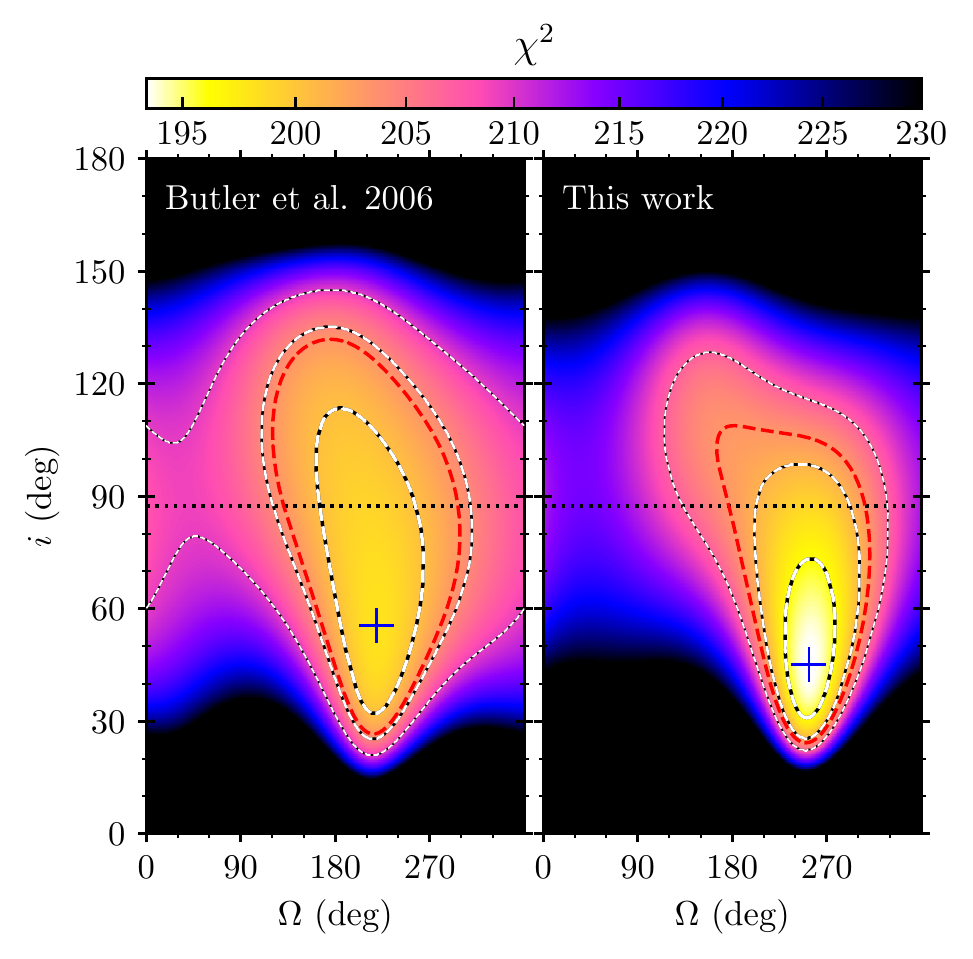}
    \caption{Goodness of fit ($\chi^2$) as a function of the inclination $i$ and the position angle of the ascending node $\Omega$ for the photocenter orbit of the $\pi$ Men system using the spectroscopic orbit from \citet{Butler:2006dd} (left), and the revised orbit presented in Table~\ref{tbl:rv-ml} (right). Dashed lines (black and white) denote areas enclosing 68, 95, and 99.7\,\% of the probability. The value of $\chi^2_{\rm null}$ calculated using an unperturbed five-parameter fit is also plotted (red dashed line). The lowest $\chi^2$ is indicated by the cross, and the inclination of $\pi$ Men c from \citet{Huang:2018dg} is plotted for reference (horizontal dashed line).}.
    \label{fig:chi2-hipOnly}
\end{figure}
We first applied this model to $\pi$ Men using the spectroscopic orbital elements from \cite{Butler:2006dd} both to compare with previous results from \cite{Reffert:2011ca} and to demonstrate the effect of a revised spectroscopic orbit on the determination of the astrometric orbit of the photocenter. With data spanning slightly more than one orbital period, \citet{Butler:2006dd} measured $P=2151$\,d, $K_1=196.4$\,m\,s$^{-1}$, $e=0.6405$, $\omega_\star=330\fdg24$ and $T_0=47819.5$ MJD. We used these spectroscopic elements and their estimate for the mass of the primary of $1.1$\,$M_{\odot}$. We used the same size and range for the $i$-$\Omega$ grid described previously. The small subset of inclinations that correspond to orbital configurations where $M_2>M_1$ are ignored by setting $\chi^2=\infty$.

The resulting $\chi^2$ surface is shown in Figure~\ref{fig:chi2-hipOnly} (left panel). We found a minimum $\chi^2$ of $198.0$ at $i=55\fdg6$ and $\Omega=219\fdg5$ ($\chi^2_{\nu}=1.50$ given 137 measurements and five free parameters), with marginalized $1\sigma$ confidence intervals of $49\fdg6$--$99\fdg2$ for $i$ and $180\fdg5$--$248\fdg6$ for $\Omega$. The minimum $\chi^2$ is a slight improvement over the single star five-parameter astrometric model ($\chi^2_{\rm null}=203.2$, $\chi^2_{\nu, {\rm null}} = 1.54$), however the reduced $\chi^2$ in both cases suggesting a slight underestimate in the {\it Hipparcos} abscissa uncertainties. Our results are consistent with the $3\sigma$ joint confidence interval on the inclination in \cite{Reffert:2011ca}; we found a range of $21\fdg4$--$144\fdg6$, compared to $20\fdg3$--$150\fdg6$ within their study.

This technique is dependent on having a good estimate of the motion of the photocenter of the system during time span of the {\it Hipparcos} observations. The radial velocity record used by \citet{Butler:2006dd} covered one and a third orbital periods, but only sampled the part of the orbit with the maximum velocity change near periastron once. This resulted in quite a large uncertainty in the period ($\sigma_P=85$\,d) which translated into an uncertainty of the epoch of periastron in late 1989/early 1990 of $\sigma_{T_0}=170$\,d. This causes significant uncertainty in the model of the astrometric motion of the photocenter for a given $M_1$, $i$, and $\Omega$.

We used the radial velocity record that we compiled (Section~\ref{sec:rv}) to revise the measurements of the spectroscopic orbital elements. We used maximum likelihood estimation to determine the Keplerian orbit that best fit the data. The radial velocity of the star $v$ is calculated from the spectroscopic orbital elements as
\begin{equation}
    v = K_1\left[e\cos\omega_\star+ \cos\left(\nu + \omega_\star\right)\right] + \gamma.
\end{equation}
Here $\gamma$ is the time-dependent apparent radial velocity of the system barycenter; secular acceleration due to the changing perspective between us and the $\pi$ Mensae system cause a significant change in the apparent radial velocity of the barycenter, which is assumed to be moving linearly through space, over the twenty-year radial velocity record. We used the coordinate transformations described in \citet{Butkevich:2014jt} and the {\it Hipparcos} astrometry in Table~\ref{tbl:astro} to account for this effect (see also Section \ref{sec:gaia-model}). We evaluated the likelihood $\mathcal{L}$ using the predicted velocities derived from of a given set of orbital parameters as
\begin{equation}
\begin{split}
    \ln \mathcal{L}_i = -\frac{1}{2} \sum_j^{n_i}\bigg[&\frac{\left(v_j^{\rm model} - \left[v_j^{\rm obs}+\Delta v_i\right]\right)^2}{\sigma^2_j + \epsilon_i^2}\\
    & + \ln \left(2\pi\left[\sigma^2_j + \epsilon^2_i\right]\right)\bigg]
\end{split}
\end{equation}
where $v_j^{\rm model}$ and $v_j^{\rm obs}$ are the predicted and measured velocities for the j$^{\rm th}$ epoch, and $\sigma_j$ is the uncertainty of the measurement. The likelihood is evaluated for each instrument $i$ independently. The summation is performed over the $n_i$ measurements for the $i^{\rm th}$ instrument, with two additional terms describing the radial velocity offset for that instrument $\Delta v_i$, and an error inflation term $\epsilon_i$ to account for both underestimated uncertainties and stellar jitter (e.g., \citealp{PriceWhelan:2017br,Fulton:2018gq}). The final likelihood is then simply $\ln\mathcal{L} = \sum_i\ln\mathcal{L}_i$. We used the HARPS measurements taken prior to the replacement of the fibres (``HARPS1'') to define our absolute radial velocity zero-point by fixing $\Delta v_i$ to zero for this instrument.

\begin{figure*}
    \centering
    \includegraphics[width=\textwidth]{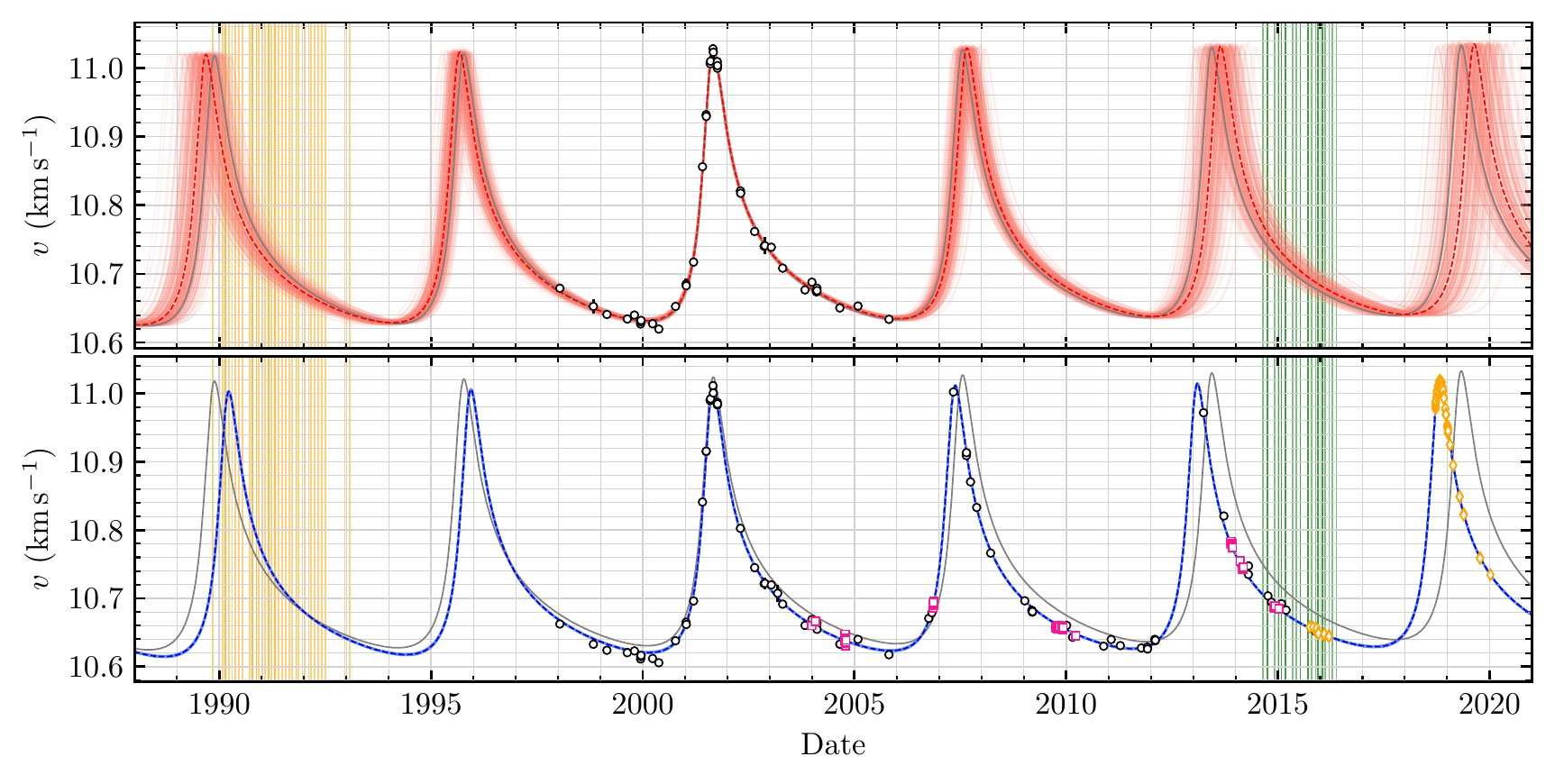}
    \caption{Spectroscopic orbit of the star $\pi$ Men around the system barycenter using the radial velocity record presented in \citet{Butler:2006dd} (top panel) and this work (bottom panel). Measurements are plotted from UCLES (black circles), HARPS1 (pre-upgrade, pink squares), and HARPS2 (post-upgrade; yellow diamonds). The maximum likelihood orbit is plotted as a dashed line, with draws from an MCMC fit shown as light solid lines. The orbit is now so well constrained that the uncertainty of the predicted radial velocity at any given epoch is smaller than the thickness of the dashed line. In both panels the best fit orbit computed by \citet{Butler:2006dd} and used by \citet{Reffert:2011ca} is shown as a solid gray curve. Dates of the individual {\it Hipparcos} and {\it Gaia} measurements are denoted by orange and green vertical lines, respectively.} 
    \label{fig:rv}
\end{figure*}
\begin{table}
\caption{Maximum likelihood estimate of the spectroscopic orbit of $\pi$ Mensae b}
\label{tbl:rv-ml}
 \centering
\begin{tabular}{lcc}
\hline\hline
Property & Value & Unit \\
\hline
$P$ & $2089.14$ & d\\
$K_1$ & $193.71$ & m\,s$^{-1}$\\
$\sqrt{e}\cos\omega_\star$ & $0.703$ & -\\
$\sqrt{e}\sin\omega_\star$ & $-0.388$ & -\\
$T_0$ & $52123.12$ & MJD\\
$\gamma_{1991.25}$ & $10700.19$ & m\,s$^{-1}$\\
$\Delta v_{\rm H1}$ & $\equiv 0$ & m\,s$^{-1}$\\
$\Delta v_{\rm H2}$ & $-16.76$ & m\,s$^{-1}$\\
$\Delta v_{\rm U}$ & $10675.31$ & m\,s$^{-1}$\\
$\epsilon_{\rm H1}$ & $3.12$ & m\,s$^{-1}$\\
$\epsilon_{\rm H2}$ & $2.00$ & m\,s$^{-1}$\\
$\epsilon_{\rm U}$ & $5.53$ & m\,s$^{-1}$\\
\hline
\multicolumn{3}{c}{Derived parameters}\\
\hline
$e$ & $0.645$ & -\\
$\omega_\star$ & $331.14$ & deg\\
\hline
\end{tabular}
\end{table}

Table~\ref{tbl:rv-ml} contains the maximum likelihood estimate of the set of spectroscopic orbital elements for the orbit of $\pi$ Men b. We used a typical parameterization for the spectroscopic elements where the eccentricity $e$ and argument of periastron $\omega_{\star}$ are combined into $\sqrt{e}\sin\omega_{\star}$ and $\sqrt{e}\cos\omega_{\star}$ to avoid angle wrapping. The effect of the inner planet on these orbital parameters is negligible due to the significantly smaller velocity semi-amplitude and shorter orbital period. We repeat the analysis of the {\it Hipparcos} IAD using this updated set of orbital elements. Here we used the mass estimate of $M_1=1.094\pm0.039$\,$M_{\odot}$ from \cite{Huang:2018dg}. The resulting $\chi^2$ surface is shown in Figure~\ref{fig:chi2-hipOnly} (right panel), showing a significant improvement of the constraint of the inclination of the orbit. We found a minimum $\chi^2$ of 193.4 at $i=45\fdg1$ and $\Omega=253\fdg0$ ($\chi^2_\nu=1.47$), an improved goodness of fit with the more accurate model of the photocenter motion during the time span of the {\it Hipparcos} measurements. The marginalized $1\sigma$ confidence intervals for $i$ and $\Omega$ are $37\fdg6$--$66\fdg0$ and $237\fdg5$--$269\fdg2$. An edge-on orbit of $i\sim90$\,deg for $\pi$ Men b is weakly excluded when only using the {\it Hipparcos} IAD at the $\sim1.5\sigma$ level.

\section{Incorporating Gaia astrometry}
\label{sec:gaia}
Astrometric measurements of bright stars such as $\pi$ Men with {\it Gaia} have a typical per-scan measurement error of $\sim$1\,mas, comparable to that achieved with {\it Hipparcos}, and worse than predicted from models of the instrument \citep{Lindegren:2018gy}. Despite this, the combination of two extremely precise measurements of the photocenter location separated by approximately 24 years, the {\it Hipparcos} IAD, and the instantaneous proper motion measured by {\it Gaia} can be used to measure deviations from linear motion caused by the presence of an orbiting companion (e.g., \citealp{Kervella:2019bw}).

Without access to the individual astrometric measurements used to fit the astrometric parameters given within the catalogue it is not possible to perform the same analysis as applied to the {\it Hipparcos} data described in Section~\ref{sec:hip-only}. Instead, we used the predicted scan timings, orientations, and parallax factors to forward model the {\it Gaia} measurement of the position and motion of the photocenter at the {\it Gaia} reference epoch of 2015.5. This method will be most reliable with a relatively constant photocenter motion over the time span of the {\it Gaia} scans of the star, with only a small deviation from linear motion. The {\it Gaia} scans of $\pi$ Men are sampling a relatively slow part of the orbit (Figure~\ref{fig:rv}, bottom panel), with a constant rate of astrometric acceleration of $\sim$1\,mas\,yr$^{-2}$ predicted from the best fit orbit from Section~\ref{sec:hip-only}, depending on the orbital inclination. This is comparable in magnitude to the very smallest accelerations detected by {\it Hipparcos} with a longer baseline of observations; 99\% of the stars where the accelerating model was a better fit than the linear motion model had accelerations of $>1.6$\,mas\,yr$^{-2}$. The relatively constant motion of the photocenter also mitigates the effect of the discrepancy between the number of predicted scans (26) and the actual number of scans used in the {\it Gaia} catalogue (21) as described in Section~\ref{sec:acq-astro}, especially if these five missing scans are randomly distributed throughout the time span of the {\it Gaia} measurements. The upcoming {\it Gaia} data releases will contain measurements of astrometric accelerations where appropriate with which it will be possible for a more precise simulation of the {\it Gaia} measurement without access to the individual measurements themselves.

\subsection{Model description}
\label{sec:gaia-model}
\begin{table*}
\caption{Joint astrometric-spectroscopic model parameters}
\label{tbl:params}
 \centering
\begin{tabular}{lclc}
\hline\hline
Property & Unit & Description & Prior \\
\hline
\multicolumn{4}{c}{Spectroscopic parameters}\\
\hline
$P$ & d & Orbital period & $\mathcal{U}(1, 90000)$\\
$K_1$ & m\,s$^{-1}$ & Radial velocity semi-amplitude & $\mathcal{U}(0,1{\rm \,km\,s}^{-1})$\\
$\sqrt{e}\cos\omega_\star$ & - & Paramaterized variable of eccentricity and argument of periastron & $\mathcal{U}(-1,1)$\tablefootmark{a}\\
$\sqrt{e}\sin\omega_\star$ & - & Paramaterized variable of eccentricity and argument of periastron & $\mathcal{U}(-1,1)$\tablefootmark{a}\\
$\tau$ & - & Mean anomaly at reference epoch (51000 MJD) in fractions of $P$ & $\mathcal{U}[0, 1)$\\
$\gamma_{1991.25}$ & m\,s$^{-1}$ & Radial velocity of $\pi$ Mensae system barycenter at 1991.25 (48348.25 MJD) & $\mathcal{U}(-100, 100)$\\
$\Delta v_{\rm H2}$ & m\,s$^{-1}$ & Radial velocity offset for post-upgrade HARPS & $\mathcal{U}(-100, 100)$\\
$\Delta v_{\rm U}$ & m\,s$^{-1}$ & Radial velocity offset for UCLES & $\mathcal{U}(-100, 100)$\\
$\epsilon_{\rm H1}$ & m\,s$^{-1}$ & Error inflation for HARPS1 & $\mathcal{U}(0, 1000)$\tablefootmark{b}\\
$\epsilon_{\rm H2}$ & m\,s$^{-1}$ & Error inflation for HARPS2 & $\mathcal{U}(0, 1000)$\tablefootmark{b}\\
$\epsilon_{\rm U}$ & m\,s$^{-1}$ & Error inflation for UCLES & $\mathcal{U}(0, 1000)$\tablefootmark{b}\\
$\epsilon_{\rm HIP}$ & mas & Error inflation for {\it Hipparcos} abscissa measurements & $\mathcal{U}(0, 10)$\tablefootmark{b}\\
\hline
\multicolumn{4}{c}{Astrometric parameters}\\
\hline
$\cos i$ & - & Cosine of the orbital inclination & $\mathcal{U}[-1, 1]$\\
$\Omega$ & deg & Position angle of the ascending node & $\mathcal{U}[0, 360)$\\
$M_1$ & $M_{\odot}$ & Mass of primary star & $\mathcal{N}(1.094, 0.039^2)$\\
$\Delta\alpha^\star$ & mas & R.A. offset between {\it Hipparcos} measurement and barycenter at 1991.25 & $\mathcal{U}(-20, 20)$ \\
$\Delta\delta$ & mas & Dec. offset -----"----- & $\mathcal{U}(-20, 20)$ \\
$\Delta\varpi$ & mas & Parallax offset -----"-----& $\mathcal{U}(-20, 20)$ \\
$\Delta\mu_{\alpha^\star}$ & mas\,yr$^{-1}$ & R.A. proper motion offset -----"----- & $\mathcal{U}(-20, 20)$ \\
$\Delta\mu_{\delta}$ & mas\,yr$^{-1}$ & Dec. proper motion offset -----"----- & $\mathcal{U}(-20, 20)$ \\
\hline
\end{tabular}
\tablefoot{
\tablefoottext{b}{An additional criterion of $0\le e<1$ was applied.}\\
\tablefoottext{a}{A penalty term is added to the likelihood to further constrain the possible values of these error inflation terms.}}
\end{table*}
The model was adapted from the one used in \cite{DeRosa:2019iq} to handle radial velocity data from multiple instruments and the lack of of any relevant direct imaging constraints for this system. The model consists of twenty parameters that allow us to connect our physical model of the system to the astrometric and spectroscopic observations. These parameters are described in Table~\ref{tbl:params}. In this model the barycenter of the $\pi$ Men system in 1991.25 is located at
\begin{equation}
    \alpha_0 = \alpha_{\rm H} + \Delta \alpha^\star,\; \delta_0 = \delta_{\rm H} + \Delta \delta,
\end{equation}
with a parallax and proper motion of
\begin{equation}
    \varpi = \varpi_{\rm H} + \Delta\varpi, \; \mu_{\alpha^\star} = \mu_{\alpha^\star, {\rm H}} + \Delta \mu_{\alpha^\star}, \; \mu_\delta= \mu_{\delta, {\rm H}} + \Delta \mu_\delta,
\end{equation}
where the H subscript denotes values from the {\it Hipparcos} catalogue. We followed the same procedure described in Section~\ref{sec:hip-only} to generate a model {\it Hipparcos} abscissa $\Lambda$ using the parallax $\varpi$, the offsets to the remaining four astrometric parameters, and the variable $\Upsilon$ that encodes the photocenter motion for a given set of orbital parameters. The likelihood of the set of model parameters given the {\it Hipparcos} measurements was evaluated as
\begin{equation}
    \ln \mathcal{L_{\rm H}} = -\frac{1}{2}\sum\left[\frac{\left(\Lambda - \Lambda_{\rm HIP}\right)^2}{\sigma_\Lambda^2 + \epsilon_{\rm HIP}^2} + \ln \left(2\pi\left[\sigma_{\Lambda}^2 + \epsilon_{\rm HIP}^2\right]\right)\right].
\end{equation}
The error inflation term $\epsilon_{\rm HIP}$ was added to account for the apparent underestimation of the {\it Hipparcos} IAD uncertainties noted in Section~\ref{sec:rv-fit}.

The $\pi$ Men barycenter was then propagated from the {\it Hipparcos} reference epoch (J1991.25) to the {\it Gaia} reference epoch (J2015.5) using the rigorous coordinate transformation procedure described in \citet{Butkevich:2014jt}. This procedure propagates the spherical coordinates, parallax, and tangent-plane proper motions, accounting for the non-rectilinear nature of the ICRS coordinate system. This transformation is critical given the declination of the star; applying the simplistic transformation using the tangent plane approximation would result in a 2.9\,mas error in the position and a 0.2\,mas\,yr$^{-1}$ error in the proper motion of the star at the {\it Gaia} reference epoch, mimicking photocenter motion due to the orbiting companion. This coordinate transform also accounts for the secular acceleration of the radial velocity of the system barycenter. As in Section~\ref{sec:rv-fit}, we fit for the apparent radial velocity of the system barycenter at 1991.25 ($\gamma_{1991.25}$).

With the barycenter propagated to 2015.5, we simulated the motion of the photocenter during the time span of the {\it Gaia} measurements using the propagated parallax and proper motions, and the predicted photocenter motion caused by the orbiting companion. The photocenter orbit semi-major axis was calculated using $\beta_G$ due to the differing filter responses of {\it Hipparcos} and {\it Gaia}. We then performed a simple least-squares fit to the motion of the star to simulate the {\it Gaia} measurement of the five astrometric parameters presented in the catalogue using the scan timings, angles, and parallax factors obtained from the {\tt gost} utility described previously. We constructed a residual vector $\vec{r}$ containing the difference between the catalogue and simulated position and proper motion measurements. This was used to calculate the likelihood of the model parameters given the {\it Gaia} measurements as
\begin{equation}
    \ln \mathcal{L}_{\rm G} = -\frac{1}{2}\vec{r}^T\mathbf{C}^{-1}\vec{r}
\end{equation}
where $\mathbf{C}$ is the covariance matrix of the {\it Gaia} catalogue measurements constructed using the uncertainties and relevant correlation coefficients.

The likelihood of the model parameters given the radial velocity measurements was calculated as in Section~\ref{sec:rv-fit}
\begin{equation}
    \ln \mathcal{L}_{\rm RV} = \sum_i^n\ln\mathcal{L}_i
\end{equation}
for each of the $n$ instruments used to construct the radial velocity record in Table~\ref{tbl:rvs}. The final likelihood was calculated by summing the three components
\begin{equation}
    \label{eq:lnlike}
    \ln\mathcal{L} = \ln\mathcal{L}_{\rm H} + \ln\mathcal{L}_{\rm G} + \ln\mathcal{L}_{\rm RV}
\end{equation}

\subsection{Application to $\pi$ Mensae}
\begin{figure}
    \centering
    \includegraphics[width=0.5\textwidth]{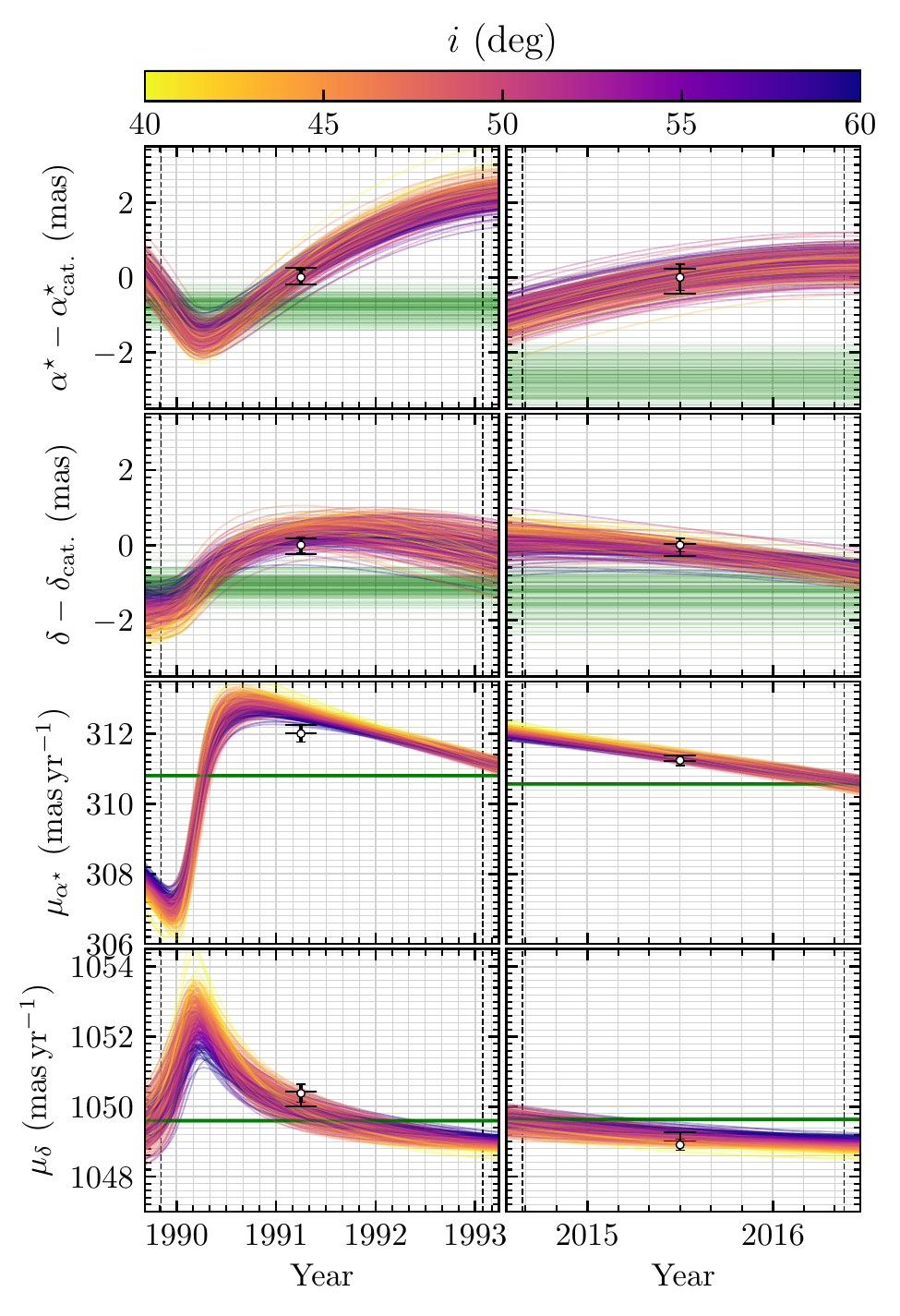}
    \caption{Position (top two panels) and proper motion (bottom two panels) of the photocenter during time span of the {\it Hipparcos} and {\it Gaia} measurements used in this study from the fit using all available astrometric data. For the position plots the proper motion of the barycenter was subtracted, and the positions are relative to the catalogue values. The position of the barycenter relative to the catalogue value and its proper motion are also plotted (horizontal green lines). The significant change in the proper motion of the barycenter in the $\alpha^{\star}$ direction is due to the southern declination of the star. Catalogue values are plotted as the circle with thick error bars. Simulated measurements based on a five-parameter fit to the combined motion of the photocenter and barycenter are shown by the more extended error bars.}
    \label{fig:pm-fit}
\end{figure}
\begin{table*}
\tiny
\caption{Maximum likelihood estimate and MCMC median and $1\sigma$ credible intervals for the fitted and derived parameters}
\label{tbl:results}
\centering
\begin{tabular}{lccc|cc|cc|cc}
\hline\hline

Property & Unit & \multicolumn{2}{c}{{\it Hipparcos} only} & \multicolumn{2}{c}{{\it Hipparcos}, {\it Gaia} pos.}  & \multicolumn{2}{c}{{\it Hipparcos}, {\it Gaia} p.m.}  & \multicolumn{2}{c}{{\it Hipparcos}, {\it Gaia} pos. \& p.m.}\\
 & & $\mathcal{L}_{\rm max}$ & MCMC &$\mathcal{L}_{\rm max}$ & MCMC &$\mathcal{L}_{\rm max}$ & MCMC &$\mathcal{L}_{\rm max}$ & MCMC\\
\hline
\multicolumn{10}{c}{Spectroscopic parameters}\\
\hline
$P$ & d & $2089.14$ & $2089.13_{-0.37}^{+0.36}$ & $2089.14$ & $2089.14_{-0.37}^{+0.36}$ & $2089.14$ & $2089.13_{-0.37}^{+0.36}$ & $2089.11$ & $2089.11_{-0.37}^{+0.36}$\\
$K_1$ & m\,s$^{-1}$ & $193.71$ & $193.71 \pm 0.33$ & $193.71$ & $193.71 \pm 0.33$ & $193.71$ & $193.71 \pm 0.33$ & $193.70$ & $193.70 \pm 0.33$\\
$\sqrt{e}\cos\omega_\star$ & - & $0.7034$ & $0.7034 \pm 0.0014$ & $0.7034$ & $0.7034 \pm 0.0014$ & $0.7034$ & $0.7034 \pm 0.0014$ & $0.7034$ & $0.7034 \pm 0.0014$\\
$\sqrt{e}\sin\omega_\star$ & - & $-0.3877$ & $-0.3877_{-0.0030}^{+0.0031}$ & $-0.3877$ & $-0.3877_{-0.0030}^{+0.0031}$ & $-0.3877$ & $-0.3876_{-0.0030}^{+0.0031}$ & $-0.3876$ & $-0.3875_{-0.0030}^{+0.0031}$\\
$\tau$ & - & $0.53760$ & $0.53760_{-0.00065}^{+0.00067}$ & $0.53760$ & $0.53760_{-0.00065}^{+0.00067}$ & $0.53760$ & $0.53761_{-0.00065}^{+0.00066}$ & $0.53764$ & $0.53765_{-0.00065}^{+0.00067}$\\
$\gamma_{1991.25}$ & m\,s$^{-1}$ & $10700.25$ & $10700.24 \pm 0.33$ & $10700.24$ & $10700.25 \pm 0.33$ & $10700.25$ & $10700.25 \pm 0.33$ & $10700.23$ & $10700.23 \pm 0.33$\\
$\Delta v_{\rm H2}$ & m\,s$^{-1}$ & $-16.79$ & $-16.79_{-0.50}^{+0.49}$ & $-16.79$ & $-16.79_{-0.49}^{+0.50}$ & $-16.79$ & $-16.79_{-0.49}^{+0.50}$ & $-16.78$ & $-16.79_{-0.49}^{+0.50}$\\
$\Delta v_{\rm U}$ & m\,s$^{-1}$ & $10675.32$ & $10675.32_{-0.76}^{+0.77}$ & $10675.32$ & $10675.32 \pm 0.76$ & $10675.32$ & $10675.32 \pm 0.76$ & $10675.32$ & $10675.32 \pm 0.76$\\
$\epsilon_{\rm H1}$ & m\,s$^{-1}$ & $3.12$ & $3.17_{-0.21}^{+0.23}$ & $3.12$ & $3.17_{-0.21}^{+0.23}$ & $3.12$ & $3.17_{-0.21}^{+0.23}$ & $3.12$ & $3.17_{-0.21}^{+0.23}$\\
$\epsilon_{\rm H2}$ & m\,s$^{-1}$ & $2.00$ & $2.04_{-0.12}^{+0.13}$ & $2.00$ & $2.04_{-0.12}^{+0.13}$ & $2.00$ & $2.04_{-0.12}^{+0.13}$ & $2.00$ & $2.04_{-0.12}^{+0.13}$\\
$\epsilon_{\rm U}$ & m\,s$^{-1}$ & $5.54$ & $5.70_{-0.51}^{+0.58}$ & $5.54$ & $5.70_{-0.51}^{+0.58}$ & $5.54$ & $5.70_{-0.51}^{+0.58}$ & $5.55$ & $5.71_{-0.51}^{+0.58}$\\
$\epsilon_{\rm HIP}$ & mas & $0.83$ & $0.94_{-0.21}^{+0.20}$ & $0.84$ & $0.91_{-0.21}^{+0.20}$ & $0.84$ & $0.91_{-0.21}^{+0.20}$ & $0.87$ & $0.92_{-0.21}^{+0.20}$\\
\hline
\multicolumn{10}{c}{Astrometric parameters}\\
\hline
$\cos i$ & - & $0.70$ & $0.41_{-0.43}^{+0.27}$ & $0.56$ & $0.48_{-0.25}^{+0.14}$ & $0.72$ & $0.69_{-0.09}^{+0.07}$ & $0.66$ & $0.64_{-0.07}^{+0.06}$\\
$\Omega$ & deg & $253.5$ & $250.7_{-33.5}^{+25.0}$ & $252.7$ & $248.6_{-15.8}^{+13.4}$ & $255.1$ & $254.8_{-10.3}^{+10.2}$ & $269.8$ & $270.3_{-8.0}^{+8.1}$\\
$M_1$ & $M_\odot$ & $1.094$ & $1.094 \pm 0.039$ & $1.094$ & $1.094 \pm 0.039$ & $1.094$ & $1.094 \pm 0.039$ & $1.094$ & $1.094 \pm 0.039$\\
$\Delta\alpha^\star$ & mas & $-0.51$ & $-0.50_{-0.35}^{+0.45}$ & $-0.48$ & $-0.49 \pm 0.28$ & $-0.55$ & $-0.54_{-0.30}^{+0.31}$ & $-0.78$ & $-0.78_{-0.26}^{+0.25}$\\
$\Delta\delta$ & mas & $-1.48$ & $-0.81_{-0.65}^{+0.59}$ & $-1.08$ & $-0.94_{-0.31}^{+0.33}$ & $-1.55$ & $-1.44 \pm 0.32$ & $-1.12$ & $-1.07 \pm 0.27$\\
$\Delta\varpi$ & mas & $0.23$ & $0.21_{-0.25}^{+0.24}$ & $0.22$ & $0.23 \pm 0.23$ & $0.22$ & $0.22 \pm 0.24$ & $0.08$ & $0.10 \pm 0.23$\\
$\Delta\mu_{\alpha^\star}$ & mas\,yr$^{-1}$ & $-1.37$ & $-1.16_{-0.39}^{+0.53}$ & $-1.18$ & $-1.18 \pm 0.02$ & $-1.45$ & $-1.38 \pm 0.18$ & $-1.20$ & $-1.20 \pm 0.02$\\
$\Delta\mu_\delta$ & mas\,yr$^{-1}$ & $-1.11$ & $-0.70_{-0.56}^{+0.67}$ & $-0.80$ & $-0.80 \pm 0.02$ & $-1.15$ & $-1.16 \pm 0.21$ & $-0.79$ & $-0.78 \pm 0.02$\\
\hline
\multicolumn{10}{c}{Derived parameters}\\
\hline
$a$ & au & $3.309$ & $3.307 \pm 0.039$ & $3.307$ & $3.307 \pm 0.039$ & $3.309$ & $3.309 \pm 0.039$ & $3.309$ & $3.308 \pm 0.039$\\
$i$ & deg & $45.9$ & $65.5_{-18.9}^{+25.5}$ & $55.8$ & $61.4_{-9.5}^{+15.6}$ & $44.2$ & $46.6_{-5.5}^{+7.0}$ & $48.7$ & $49.9_{-4.5}^{+5.3}$\\
$e$ & - & $0.6451$ & $0.6450 \pm 0.0011$ & $0.6451$ & $0.6450 \pm 0.0011$ & $0.6451$ & $0.6450 \pm 0.0011$ & $0.6450$ & $0.6450 \pm 0.0011$\\
$\omega_\star$ & deg & $331.14$ & $331.14 \pm 0.24$ & $331.14$ & $331.14 \pm 0.24$ & $331.14$ & $331.14 \pm 0.24$ & $331.15$ & $331.15_{-0.23}^{+0.24}$\\
$T_0$ & MJD & $52123.11$ & $52123.12_{-1.17}^{+1.20}$ & $52123.11$ & $52123.12_{-1.17}^{+1.20}$ & $52123.12$ & $52123.14_{-1.18}^{+1.20}$ & $52123.19$ & $52123.21_{-1.18}^{+1.20}$\\
$M_2$ & $M_{\rm Jup}$ & $13.87$ & $11.07_{-0.99}^{+2.68}$ & $12.04$ & $11.35_{-1.05}^{+1.34}$ & $14.29$ & $13.71_{-1.35}^{+1.49}$ & $13.26$ & $13.01_{-0.95}^{+1.03}$\\
\hline
\multicolumn{10}{c}{Goodness of fit}\\
\hline
$\chi^2_{\rm RV}$ & - & $360.7$ & - & $360.7$ & - & $360.7$ & - & $360.9$ & -\\
$\chi^2_{\rm HIP}$ & - & $143.8$ & - & $143.9$ & - & $143.8$ & - & $143.8$ & -\\
$\chi^2_{\rm Gaia-pos}$ & - & - & - & $0.00$ & - & - & - & $0.58$ & -\\
$\chi^2_{\rm Gaia-pm}$ & - & - & - & - & - & $0.00$ & - & $2.12$ & -\\
$\sum\chi^2$ & - & $504.5$ & - & $504.6$ & - & $504.5$ & - & $507.4$ & -\\
$\sum\chi^2_{\nu}$ & - & $1.06$ & - & $1.06$ & - & $1.06$ & - & $1.06$ & -\\
\hline

\end{tabular}
\end{table*}
We used a combination of maximum likelihood estimation and Markov chain Monte Carlo (MCMC) to determine the best fit model parameters and their uncertainties and covariances given the astrometric and spectroscopic measurements of the $\pi$ Men system. This was performed with four combinations of the astrometric data; {\it Hipparcos} only to compare with the results in Section~\ref{sec:hip-only}, and with the various combinations of the {\it Hipparcos} and the {\it Gaia} position and proper motion measurements. The maximum likelihood parameter set was found using the Nelder-Mead simplex algorithm within the {\tt scipy.optimize} package \citep{Virtanen:2020cp}. We fixed $M_1=1.094$\,$M_{\odot}$ for this step as this parameter is not constrained by any of the measurements. The results for each of the four combinations of astrometric data are listed in Tables~\ref{tbl:results}.

The uncertainties on the model parameters were estimated using a Markov chain Monte Carlo algorithm. We used the parallel-tempering affine-invariant sampler {\tt emcee} \citep{ForemanMackey:2013io} to thoroughly explore the posterior distributions of each parameter. We initialized 512 MCMC chains at each of 16 ``temperatures'' (a total of 8192 chains) near the best fit solution identified via the maximum likelihood analysis described previously. The lowest temperature chains for parallel-tempering MCMC explore the posterior distribution of each parameter, whilst the highest temperature chains explore the priors. This is especially advantageous for complex likelihood volumes to ensure parameter space is fully explored.

Each chain was advanced by $10^5$ steps with every tenth step saved to disk. At each step a probability was calculated by combining the likelihood from Eqn.~\ref{eq:lnlike} with the prior probabilities given the distributions listed in Table~\ref{tbl:params}. We discarded the first fifth of the chain as ``burn-in'' where the chain positions were still a function of their initial conditions. While we did not apply a statistical check for convergence, the chains appeared well-mixed based on a visual inspection of the chain position as a function of step number, with the median and $1\sigma$ ranges of each parameter not significantly changing. This process was repeated for each of the four combinations of astrometric data. The resulting median and $1\sigma$ credible intervals are given in Table~\ref{tbl:results} for the nineteen fitted parameters, along with several derived parameters. A visualization of the fit using all of the available astrometric data is shown in  Figure~\ref{fig:pm-fit}. We found $\chi^2_{\nu}=1.06$ for the maximum likelihood estimate of the fit using all of the astrometry available given the 500 measurements and 20 model parameters, but this parameters is dominated by the goodness of fit to the radial velocity measurements and the {\it Hipparcos} IAD. The quality of the fit to the {\it Gaia} measurements is good, although there is some tension between the prediction and measurement of the photocenter proper motion in the declination direction (Fig.~\ref{fig:pm-fit}). A potential source of this discrepancy is a residual systematic rotation of the bright star reference frame not corrected using the model from \citet{Lindegren:2019iv,Lindegren:2020ik}.

\subsection{Independent analysis}
In order to further verify these results, we perform a second analysis with an independent fitting routine for joint astrometry and RV fits, as previously used in ~\citet{nielsen:2020} for the directly-imaged planet $\beta$~Pictoris~b.  This routine is modified for $\pi$~Men~b, fitting for 20 parameters, including the standard set of seven for visual orbits: semi-major axis ($a$), eccentricity ($e$), inclination angle ($i$), argument of periastron of the planet ($\omega$), position angle of the ascending ($\Omega$), epoch of periastron passage ($T_0$), and period ($P$).  Also fitted are seven radial velocity parameters, including system radial velocity at 1991.25 ($\gamma_{1991.25}$), $m \sin{i}$ of the secondary, RV offsets ($\Delta v_{\rm H2}$, $\Delta v_{\rm U}$), and jitter terms for each instrument, $\epsilon_{\rm H1}$, $\epsilon_{\rm H2}$, and $\epsilon_{\rm U}$, as well as the error inflation term for the {\it Hipparcos} abscissa data $\epsilon_{\rm HIP}$.  Finally, we have five astrometric parameters; the offset of the photocenter from the {\it Hipparcos} catalog position of $\pi$ Men at 1991.25 ($\Delta \alpha^\star$, $\Delta \delta$), parallax ($\varpi$), and proper motion at 1991.25 ($\mu_{\alpha^\star}$, $\mu_\delta$).  We assume priors that are uniform in $\log{a}$, $\log{P}$, $\cos{i}$, and uniform in all other parameters, except for an additional prior on the derived mass of the primary ($M_1$), as above taken to be a Gaussian with mean of 1.094 M$_\odot$ and $\sigma$ of 0.039\,$M_\odot$.  As above, we utilize the full RV dataset for $\pi$ Men, and the {\it Gaia} DR2 position and proper motion measurements and errors.  However, we use a separate method to convert the \textit{Hipparcos} IAD residuals and scan directions into individual measurements, as in \citet{nielsen:2020}.  Unlike \citet{nielsen:2020}, we use rigorous propagation of position and proper motion from 1991.25 to 2015.5, by transforming position, proper motion, radial velocity, and parallax from angular coordinate to Cartesian, update all six values assuming linear motion, then transform back to angular coordinates.  Fitting is then done using a Metropolis-Hastings MCMC routine.

Despite the different method we find very similar results to the {\it Hipparcos}, {\it Gaia}, position and proper motion fit given in Table~\ref{tbl:results}.  This second method finds value of [semi-major axis, period, eccentricity, position angle of nodes, argument of periastron of the star, secondary mass,  inclination angle] of 
[$3.305^{+0.038}_{-0.039}$\,au, 
$2089.08^{+0.38}_{-0.38}$\,, 
$0.6438 \pm 0.0015$, 
$274.8^{+8.0\circ}_{-7.9}$, 
$330.58 \pm 0.31^\circ$, 
$12.87^{+1.04}_{-0.96}$\,$M_{\textrm{Jup}}$, 
$50.6^{+5.7 \circ}_{-4.8}$], 
compared to the values above of 
[$3.308\pm0.039$\,au,
$2089.11^{+0.37}_{-0.36}$\,d, 
$0.6450 \pm 0.0011$, 
$270.3^{+8.1 \circ}_{-8.0}$,
$331.15^{+0.24\circ}_{-0.23}$,
$13.01^{+1.03}_{-0.95}$\,$M_{\textrm{Jup}}$,
$49.9^{+5.3 \circ}_{-4.5}$]
.  For all parameters, the medians and confidence intervals are almost identical from the two methods.  This gives us further confidence in the  measurement of the parameter most of interest, inclination angle, which is essentially the same for the two orbit-fitting methods.

\subsection{Mutual inclination}
\begin{figure}
    \centering
    \includegraphics[width=0.5\textwidth]{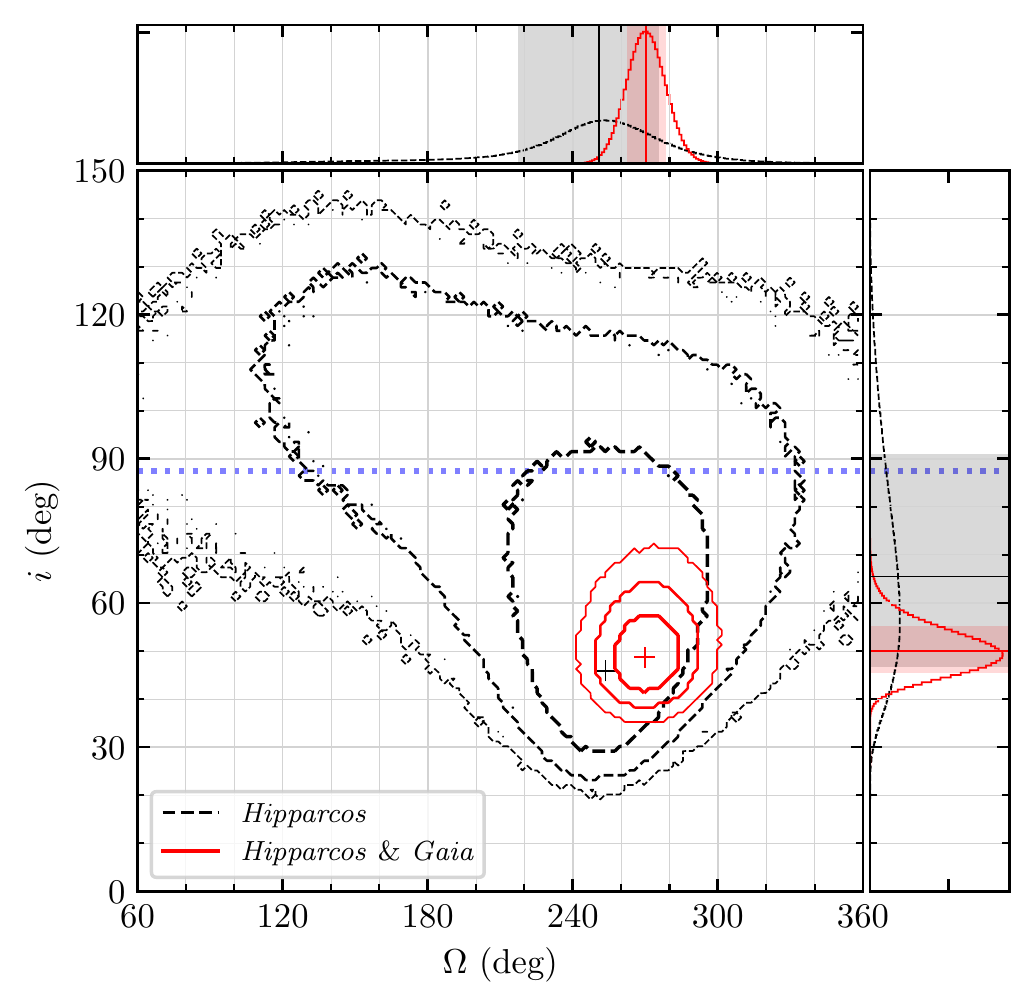}
    \caption{Covariance between the position angle of the ascending node $\Omega$ and inclination $i$, and corresponding marginalized distributions, of the photocenter orbit of the $\pi$ Men system from the MCMC calculation using only the {\it Hipparcos} measurements (black) and using all available astrometry (red). Contours denote areas encompassing 68, 95, and 99.7\,\% of the probability. Marginalized $1\sigma$ credible intervals are indicated. The values of $i$ and $\Omega$ from the maximum likelihood analyses are shown for reference (cross symbols). The orbital inclination of $\pi$~Men~b is significantly different from that of the inner planet (blue dotted line), indicating that the system is not aligned.}
    \label{fig:io-covariance}
\end{figure}
\begin{figure}
    \centering
    \includegraphics[width=0.5\textwidth]{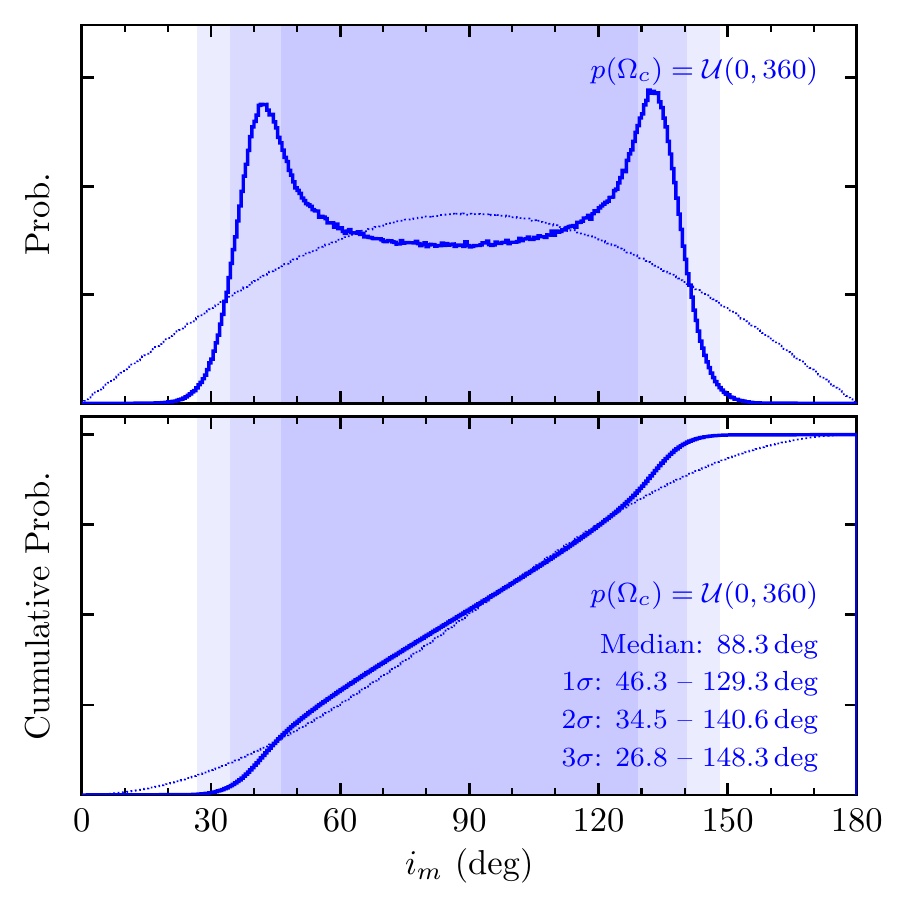}
    \caption{Posterior probability (top) and cumulative (bottom) distribution for the value of $i_{\rm mut}$ under the assumption that $p(\Omega_c) = \mathcal{U}(0, 360\,{\rm deg})$. The prior probability distribution is plotted for reference (dashed histogram).}
    \label{fig:inclination}
\end{figure}
The joint fit of the spectroscopic and all of the astrometric measurements yielded an inclination of $i=49.9_{-4.5}^{+5.3}$\,deg and a position angle of the ascending node of $\Omega=270.3_{-8.0}^{+8.1}$\,deg for the orbit of $\pi$ Men b (Fig.~\ref{fig:io-covariance}). We found consistent values for these parameters for each of the four fits shown in Table~\ref{tbl:results}. This inclination corresponds to a mass of $\pi$ Men b of $M_2=13.01_{-0.95}^{+1.03}$\,$M_{\rm Jup}$, a mass straddling the deuterium-burning limit sometimes used as the planet-brown dwarf boundary. The mutual inclination $i_{\rm mut}$ between the orbital plane of the two planets in this system cannot be directly measured as the position angle of the orbit of $\pi$ Men c ($\Omega_c$) is currently unconstrained.

We instead estimated a plausible range of the mutual inclination assuming that the position angle of the orbit of the inner planet is randomly orientated relative to that of the outer planet, $p(\Omega_c) = \mathcal{U}(0, 360\,{\rm deg})$. We estimated the posterior distribution of $i_{\rm mut}$ by combining the posterior distributions for the inclination and position angle for $\pi$ Men b ($i_b$, $\Omega_b$) calculated previously, the inclination for $\pi$ Men c ($i_c=87.456_{-0.076}^{+0.085}$\,deg; \citealp{Huang:2018dg}, although an inclination of $180-i_c$ is also consistent with the transit observations), and our assumption for the posterior distribution of $\Omega_c$;
\begin{equation}
    \cos i_{\rm mut} = \cos i_b \cos i_c + \sin i_b \sin i_c \cos\left(\Omega_b - \Omega_c\right).
\end{equation}
This yielded a 95th-percentile lower limit on the mutual inclination of $i_{\rm mut} > 34\fdg5$ given our assumption for $p(\Omega_c)$. The probability density function of the value of $i_{\rm mut}$ is shown in Figure~\ref{fig:inclination}, alongside the prior probability density function to demonstrate the constraints provided by the measurements. The minimum value of $i_{\rm mut}$ from the 4,096,000 samples was $7\fdg4$. We adopted the 95th-percentile credible interval of $i_{\rm mut}$ from $34\fdg5$ to $140\fdg6$ as the range of plausible mutual inclinations for the system given the observational data. We found a similar lower limit of $i_{\rm mut} > 30\fdg2$ when assuming $\Omega_b=\Omega_c$, indicating that the lower limit of this value is not strongly affected by the assumption made regarding the value of $\Omega_c$. A co-planar configuration for the two planets in the $\pi$ Men system is strongly excluded based on the spectroscopic and astrometric measurements used within this analysis.

\section{Dynamical constraints on present-day orbit and dynamical history}
\label{sec:dynamics}

\begin{figure}
    \centering
    \includegraphics[width=0.475\textwidth]{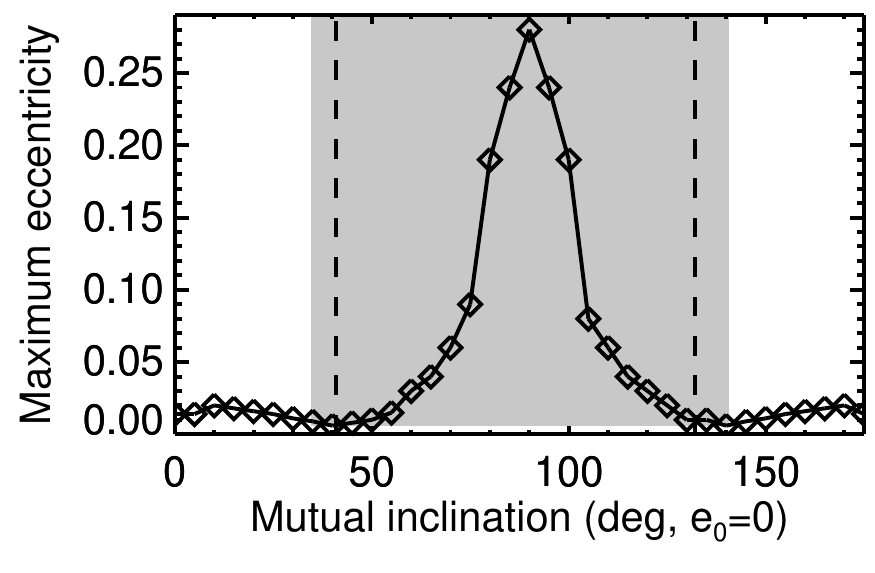}
    \caption{Maximum eccentricity of $\pi$ Men c excited by secular effects of $\pi$ Men b over 2.7 Gyr as a function of semi-major axis. We assume an initial eccentricity of 0 for $\pi$ Men c. $\pi$ Men c is shielded by general relativistic precession from strong dynamical perturbations by $\pi$ Men b. The shaded region indicates the 95\% confidence interval on the observed $i_{\rm mut}$ and the dashed lines indicate the most likely values.}
    \label{fig:maxe}
\end{figure}

\begin{figure}
    \centering
    \includegraphics[width=0.475\textwidth]{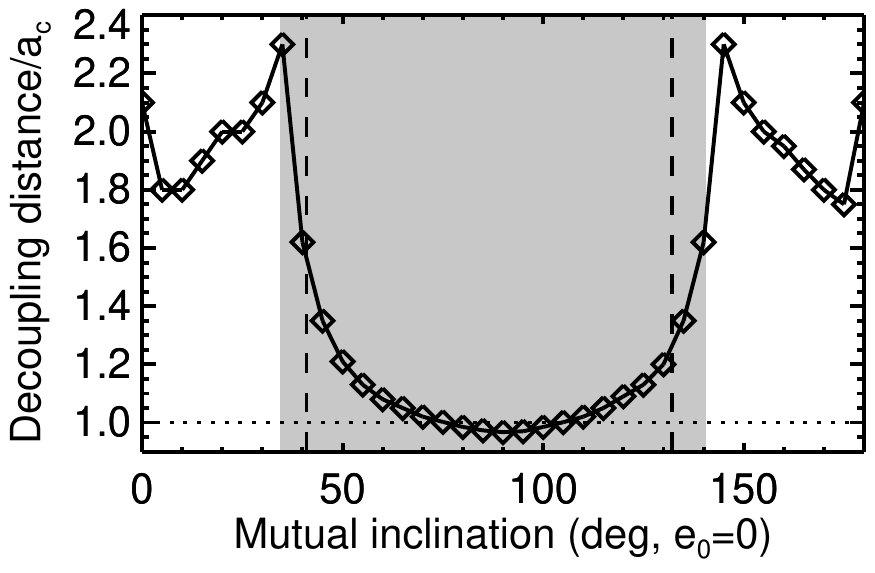}
    \caption{Decoupling distance: semi-major axis of $\pi$ Men c for which general relativistic precession prevents $\pi$ Men b from exciting its eccentricity above 0.1. We assume an initial eccentricity of 0 for $\pi$ Men c. Dotted horizontal line: $\pi$ Men c's present-day semi-major axis. The shaded region indicates the 95\% confidence interval on the observed $i_{\rm mut}$ and the dashed lines indicate the most likely values.}
    \label{fig:decoupling}
\end{figure}

On its current orbit, the inner planet $\pi$ Men c is shielded by general relativistic precession from strong dynamical perturbations by $\pi$ Men b. Therefore a long-term stability requirement does not place any additional constraints on $\pi$ Men b's orbital properties. We simulate the system at a range of mutual inclinations by approximating planet c as a test particle and solving Lagrange’s equations of motion using a secular disturbing potential expanded to octupolar order \citep{yoko03}, including general relativistic precession of both planets’ orbits. Including the octupolar order is particularly important for this nearby, eccentric outer planet (e.g., \citealt{for00,nao11,tey13,li14}). We set planet c's initial eccentricity to 0 and use the mass and orbit of planet b from Table \ref{tbl:results}. For the entire range of possible mutual inclinations, the system remains stable: planet c's eccentricity is not excited to a value high enough for tidal disruption (Fig. \ref{fig:maxe}). $\pi$ Men c's present-day semi-major axis is close to the value at which general relativistic precession prevents $\pi$ Men b from exciting its eccentricity above 0.1 (Fig. \ref{fig:decoupling}). 

$\pi$ Men c may have undergone high eccentricity tidal migration (e.g., \citealt{hut81}) to reach its present day short orbital period. In one version of this scenario, planets b and c form on or migrate to well-separated orbits. On a stable orbit, planet c must reside closer to the star than 0.69 AU \citep{petr15}. $\pi$ Men b is disturbed to its observed high eccentricity, perhaps by scattering or ejecting another planet (e.g., \citealt{jur08}). This disturbance creates a sufficient mutual inclination between planets b and c (e.g., \citealt{cha08}) to trigger Kozai-Lidov cycles (e.g., \citealt{koz62,lid62,nao16}), raising planet c’s eccentricity and triggering high eccentricity tidal migration. 

\begin{figure}
    \centering
    \includegraphics[width=0.475\textwidth]{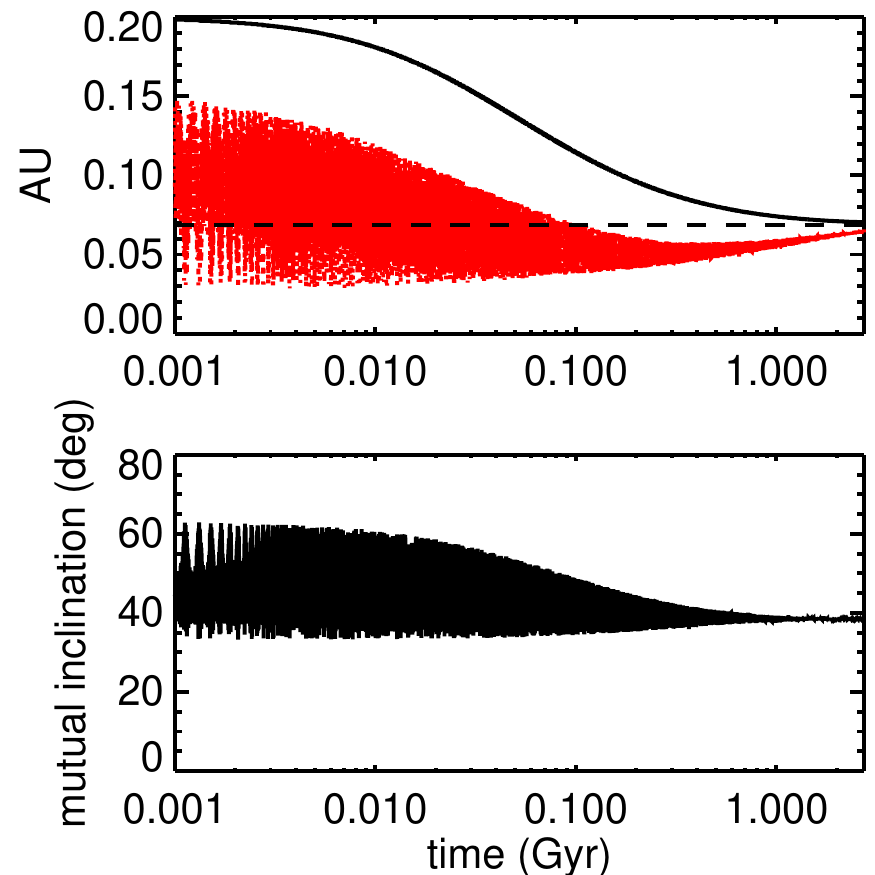}
    \caption{Example of high eccentricity Kozai-Lidov tidal migration scenario. Top: Evolution of $\pi$ Men c's semi-major axis (black) and periapse (red). Bottom panel: mutual inclination. $\pi$ Men c circularizes to its present-day semi-major axis with a mutual inclination of 38.5$^\circ$.}
    \label{fig:kozai}
\end{figure}

Figure \ref{fig:kozai} shows an example of this first scenario. We include tidal evolution following \citet{wu03} with a tidal quality factor $Q_p=10^3$ and initial mutual inclination of 45 degrees, eccentricity of 0, and semi-major of 0.2 AU for $\pi$ Men c, resulting in a mutual final inclination of 38.5$^\circ$. If planet c originally had its eccentricity excited by Kozai-Lidov cycles, underwent significant tidal migration, and is fully tidally circularized today, we expect to observe it at mutual inclination near $\sim40^\circ$ or $\sim140^\circ$ (e.g., \citealt{fab07,pet16}). Much lower present-day mutual inclinations (i.e., closer to 0 and 180) are possible if planet c began with a large eccentricity. For example, in an initially coplanar configuration where planet b has an initial eccentricity of 0.45, longitude of periapse opposite planet b, and semi-major of 0.45 AU, the final mutual inclination is 10$^\circ$. 

However, we cannot assume that planet c has completed its tidal circularization. Its observed eccentricity is not well constrained by the radial-velocity or transit light curve datasets \citep{Huang:2018dg,Gandolfi:2018cg}, and other mini-Neptunes at short orbital periods have retained their orbital eccentricities (e.g., GJ 436b, \citealt{man07}). Their tidal quality factors are not well known, with values of $Q_p$ spanning $10^2$ to $10^6$ (or an even wider range) quite plausible. We might observe a higher mutual inclination (i.e., much closer to polar than $\sim 40$ or $140^\circ$) if planet c is still in the process of tidally circularizing and has not migrated a substantial distance. For example, planet c could have started at $a=0.1$ AU and $e=0$ and mutual inclination of 65$^\circ$, been excited by Kozai-Lidov cycles onto an orbit with a periapse several times the Roche limit, experienced rather weak tidal dissipation with $Q_p=10^{7}$, and arrived to its present-day semi-major axis after 1 Gyr with $e$ oscillating between 0.6 and 0.75 and mutual inclination oscillating between 54 and 63$^\circ$ as it continues ongoing tidal evolution. At larger initial semi-major axes, near polar mutual inclinations lead to either close periapse passages that result in tidal disruption or quick tidal circularization to semi-major axes much smaller than planet c's current semi-major axis.

Another variation on the tidal migration scenario is that $\pi$ Men c underwent high eccentricity migration initially triggered by planet-planet scattering (e.g., \citealt{ras96}) or secular chaos (e.g., \citealt{wu11}) rather than planet-planet Kozai-Lidov cycles. This scenario does not require a particular mutual inclination. The evolution could involve Kozai-Lidov cycles -- or coplanar secular eccentricity cycles (e.g., \citealt{pet15}) -- once tidal evolution separates the planets. 

\begin{figure}
    \centering
    \includegraphics[width=0.475\textwidth]{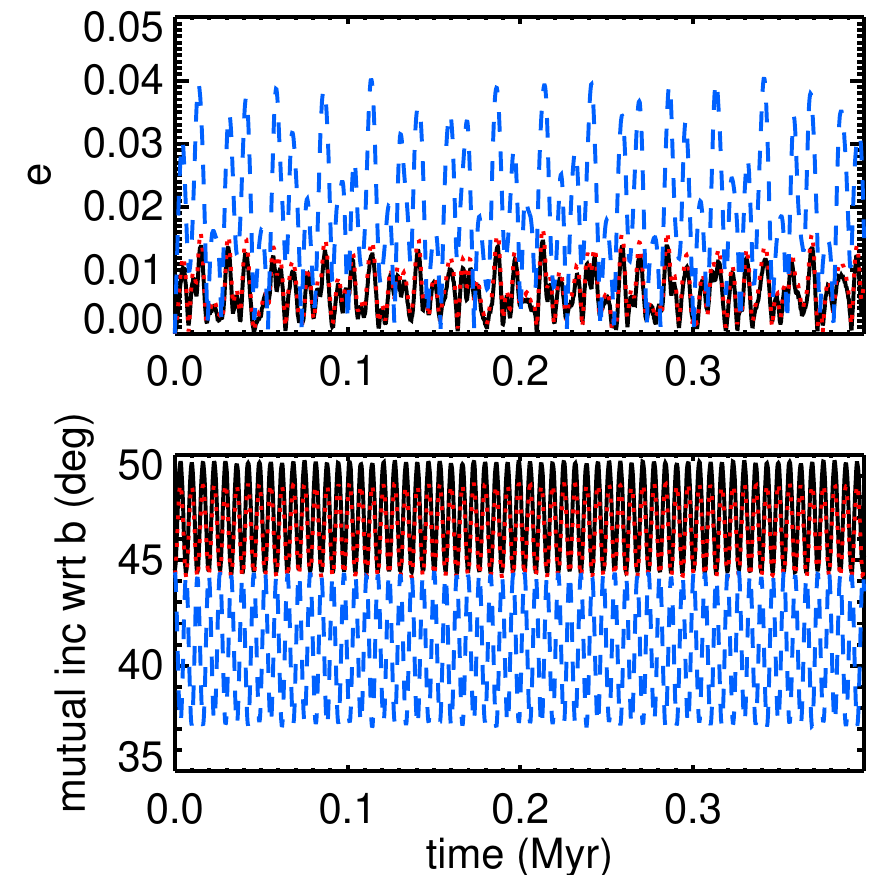}
    \caption{Example behavior of multi-planet system, with $\pi$ Men c (black solid) at its present-day semi-major axis, two other planets of the same mass located at 0.103 AU (red dotted) and 0.185 AU (blue dashed), and $\pi$ Men b. Perturbations from planet b lead to modest eccentricities (top panel) and mutual inclinations among the planets (bottom panel; i.e., each planet has a slightly different mutual inclination with planet b). Dynamical evolution computed in {\tt mercury6}) \citep{cha99}, modified to include general relativistic precession of all planets.}
    \label{fig:multi}
\end{figure}

Another possibility is that $\pi$ Men c formed at or near its current location (e.g., \citealt{lee14}), or arrived via disk migration (e.g., \citealt{coss14}). As described above, general relativistic precession mostly decouples it from planet b, with modest oscillations in eccentricity and inclination. Planet c may even reside in a close-in multi-planet system (which would not be easily compatible with the high eccentricity migration scenario above); radial velocity and transit timing constraints do not rule out the presence of other small planets. Depending on their spacing, coupling among the inner planets may cause them to oscillate in inclination together as a coplanar set, slightly excite their mutual inclinations (e.g., \citealt{lai17,hua17,masu20}; our Fig. \ref{fig:multi}), or even allow for a modest eccentricity excitation that results in short scale tidal migration of planet c. The detection of other planets in the system could allow for stronger stability constraints on mutual inclinations.

Therefore a variety of origins scenarios are consistent with the system's known properties. The significant mutual inclination makes the high eccentricity migration scenario appealing and the most likely observed mutual inclinations coincide with those expected from high eccentricity tidal migration with planet c's eccentricity originally excited by Kozai-Lidov cycles. Moreover, with a radius of 2.04 $R_\oplus$ \citep{Huang:2018dg} and period of 6.27 days, $\pi$ Men c is a member population of short period, 2--6 $R_\oplus$ planets identified by \citet{don18}, which \citet{daw18} argue are likely to have undergone high eccentricity tidal migration. However, we cannot rule out a quieter history involving {\it in situ} formation or disk migration. Crucially, this system provides the first direct measurement of a mutually inclined outer giant posited for a variety of super-Earth formation and evolution histories.

\section{Conclusion}
\label{sec:conclusion}
We have presented a joint analysis of spectroscopic and astrometric measurements of the motion of the planet hosting star $\pi$ Mensae induced by the massive ($13.01_{-0.95}^{+1.03}$\,$M_{\rm Jup}$) planet $\pi$ Mensae b orbiting with a period of 5.7 years. This analysis yielded a direct measurement of the inclination of the orbital plane of the outer planet of $i_b=49.9_{-4.5}^{+5.3}$\,deg, strongly excluding a co-planar configuration with the inner transiting super-earth. The mutual inclination between the orbital planes of the two planets is constrained to be between $34\fdg5$--$140\fdg6$ (95\% credible interval), assuming a random orientation of the position angle of the orbit of $\pi$ Men c.

Outer gas giants on mutually inclined orbits have been invoked in shaping the orbits and architectures of inner super-Earth systems, including stirring up the mutual inclinations in multi-planet systems (e.g., \citealt{hua17,lai17}) and driving super-Earths close to their stars through high eccentricity tidal migration. The $\pi$ Men system is the first directly measured example of a super-Earth accompanied by a mutually inclined giant, and is one of only three multi-planet systems with a measured and significant mutual inclination: the three super-jovians of the $\upsilon$~Andromedae system ($i_{\rm mut}=27\pm2$\,deg; \citealp{McArthur:2010jc}) and the two Saturn-mass planets of Kepler-108 ($i_{\rm mut}=24\pm15$\,deg; \citealp{Mills:2017gw}).

Histories of {\it in situ} formation, disk migration, and high eccentricity migration are all compatible with the system's current configuration. Future tighter upper limits on the inner planet's eccentricity would allow us to rule out the scenario in which $\pi$ Men c is still tidally migrating and to favor tidal eccentricity migration scenarios with a present-day mutual inclination close to 40 or 140$^\circ$ (while not ruling out {\it in situ} formation or disk migration).  Detection of nearby planets to $\pi$ Men c -- through continuing radial-velocity follow-up or transit timing variations with future TESS observations -- would favor the {\it in situ} formation or disk migration scenario, rather than long distance tidal migration, and potentially allow for tighter stability constraints on the mutual inclination between the inner planets and $\pi$ Men b. Constraints on the composition of $\pi$ Men c's atmosphere \citep{GarciaMunoz:2020dl} may provide hints to the formation location (e.g., \citealt{rog10}). However, since {\it in situ} formation could involve assembly from outer disk materials (e.g., \citealt{han12}) that could even be the in form of large icy cores, atmospheric clues to formation location are not likely to be definitive.

A single measurement of the relative astrometry between $\pi$~Men~b and its host star is likely sufficient to measure the orbital inclination without having to rely on absolute astrometric measurements such as those used in this study. A previous attempt to directly image this planet from the ground did not succeed due to the significant contrast between the host star and planet in the near-infrared, despite the observations being taken near the time of maximum angular separation \citep{Zurlo:2018da}. Developments in ground and space-based direct imaging techniques may enable a detection of this object in the near to medium-term. Indeed, the proximity of $\pi$ Men to the Earth makes it one of a handful of targets that are amenable to direct imaging in reflected light with the coronagraphic instrument (CGI; \citealp{Noecker:2016hp}) on the upcoming {\it Wide Field Infrared Survey Telescope}. The orbital elements for $\pi$~Men~b presented in this study can be used to constrain the position and expected brightness of the planet during the course of the mission, enabling observations to be timed to maximize the expected signal-to-noise ratio. Future data releases from the {\it Gaia} consortium will enable a validation of the results presented here. The upcoming DR3 will contain measurements of accelerations for sources for which a model of constant motion is a poor fit; the acceleration of the $\pi$~Men photocenter over the {\it Gaia} DR3 time span should be detectable at a significant level. The eventual release of the individual astrometric measurements made by the {\it Gaia} satellite will enable a measurement of the photocenter orbit using {\it Gaia} data alone for this, and a slew of other long-period planets detected via radial velocity.

\begin{acknowledgements}
We wish to thank both referees for their careful review that helped improve the overall quality of this work. We thank R. Wittenmyer for sharing the UCLES velocities used in this study. RID is supported by NASA Exoplanet Research Program Grant No. 80NSSC18K0355 and the Center for Exoplanets and Habitable Worlds at the Pennsylvania State University. The Center for Exoplanets and Habitable Worlds is supported by the Pennsylvania State University, the Eberly College of Science, and the Pennsylvania Space Grant Consortium. This research made use of computing facilities from Penn State's Institute for CyberScience Advanced CyberInfrastructure. ELN is supported by NSF grant No. AST-1411868 and NASA Exoplanet Research Program Grant No. NNX14AJ80G. This research has made use of the SIMBAD database, operated at CDS, Strasbourg, France \citep{Wenger:2000ef}, and the VizieR catalogue access tool, CDS, Strasbourg, France \citep{Ochsenbein:2000fm}. This work has made use of data from the European Space Agency (ESA) mission {\it Gaia} (\url{https://www.cosmos.esa.int/gaia}), processed by the {\it Gaia} Data Processing and Analysis Consortium (DPAC, \url{https://www.cosmos.esa.int/web/gaia/dpac/consortium}). Funding for the DPAC has been provided by national institutions, in particular the institutions participating in the {\it Gaia} Multilateral Agreement. 
\end{acknowledgements}

\bibliographystyle{aa}

\begin{thebibliography}{73}
\expandafter\ifx\csname natexlab\endcsname\relax\def\natexlab#1{#1}\fi

\bibitem[{Arenou {et~al.}(2018)Arenou, Luri, Babusiaux, Fabricius, Helmi,
  Muraveva, Robin, Spoto, Vallenari, Antoja, Cantat-Gaudin, Jordi, Leclerc,
  Reyl{\'e}, Romero-G{\'o}mez, Shih, Soria, Barache, Bossini, Bragaglia,
  Breddels, Fabrizio, Lambert, M~Marrese, Massari, Moitinho, Robichon,
  Ruiz-Dern, Sordo, Veljanoski, Eyer, Jasniewicz, Pancino, Soubiran, Spagna,
  Tanga, Turon, \& Zurbach}]{Arenou:2018dp}
Arenou, F., Luri, X., Babusiaux, C., {et~al.} 2018, A{\&}A, 616, A17

\bibitem[{Bryan {et~al.}(2019)Bryan, Knutson, Lee, Fulton, Batygin, Ngo, \&
  Meshkat}]{Bryan:2019em}
Bryan, M.~L., Knutson, H.~A., Lee, E.~J., {et~al.} 2019, AJ, 157, 52

\bibitem[{Butkevich \& Lindegren(2014)}]{Butkevich:2014jt}
Butkevich, A.~G. \& Lindegren, L. 2014, A{\&}A, 570, A62

\bibitem[{Butler {et~al.}(2006)Butler, Wright, Marcy, Fischer, Vogt, Tinney,
  Jones, Carter, Johnson, McCarthy, \& Penny}]{Butler:2006dd}
Butler, R.~P., Wright, J.~T., Marcy, G.~W., {et~al.} 2006, Astrophys. J., 646,
  505

\bibitem[{{Chambers}(1999)}]{cha99}
{Chambers}, J.~E. 1999, \mnras, 304, 793

\bibitem[{{Chatterjee} {et~al.}(2008){Chatterjee}, {Ford}, {Matsumura}, \&
  {Rasio}}]{cha08}
{Chatterjee}, S., {Ford}, E.~B., {Matsumura}, S., \& {Rasio}, F.~A. 2008, \apj,
  686, 580

\bibitem[{{Childs} {et~al.}(2019){Childs}, {Quintana}, {Barclay}, \&
  {Steffen}}]{chi19}
{Childs}, A.~C., {Quintana}, E., {Barclay}, T., \& {Steffen}, J.~H. 2019,
  \mnras, 485, 541

\bibitem[{{Cossou} {et~al.}(2014){Cossou}, {Raymond}, {Hersant}, \&
  {Pierens}}]{coss14}
{Cossou}, C., {Raymond}, S.~N., {Hersant}, F., \& {Pierens}, A. 2014, \aap,
  569, A56

\bibitem[{Coughlin \& L{\'o}pez-Morales(2012)}]{Coughlin:2012fz}
Coughlin, J.~L. \& L{\'o}pez-Morales, M. 2012, Astrophys. J., 750, 100

\bibitem[{{Dawson} \& {Johnson}(2018)}]{daw18}
{Dawson}, R.~I. \& {Johnson}, J.~A. 2018, \araa, 56, 175

\bibitem[{De~Rosa {et~al.}(2019)De~Rosa, Esposito, Hirsch, Nielsen, Marley,
  Kalas, Wang, \& Macintosh}]{DeRosa:2019iq}
De~Rosa, R.~J., Esposito, T.~M., Hirsch, L.~A., {et~al.} 2019, AJ, 158, 225

\bibitem[{Diego {et~al.}(1990)Diego, Charalambous, Fish, \&
  Walker}]{Diego:1990bs}
Diego, F., Charalambous, A., Fish, A.~C., \& Walker, D.~D. 1990, in Astronomy
  '90, Tucson AZ, 11-16 Feb 90, ed. D.~L. Crawford (SPIE), 562

\bibitem[{{Dong} {et~al.}(2018){Dong}, {Xie}, {Zhou}, {Zheng}, \&
  {Luo}}]{don18}
{Dong}, S., {Xie}, J.-W., {Zhou}, J.-L., {Zheng}, Z., \& {Luo}, A. 2018,
  Proceedings of the National Academy of Science, 115, 266

\bibitem[{{Fabrycky} \& {Tremaine}(2007)}]{fab07}
{Fabrycky}, D. \& {Tremaine}, S. 2007, \apj, 669, 1298

\bibitem[{{Ford} {et~al.}(2000){Ford}, {Kozinsky}, \& {Rasio}}]{for00}
{Ford}, E.~B., {Kozinsky}, B., \& {Rasio}, F.~A. 2000, \apj, 535, 385

\bibitem[{Foreman-Mackey {et~al.}(2013)Foreman-Mackey, Hogg, Lang, \&
  Goodman}]{ForemanMackey:2013io}
Foreman-Mackey, D., Hogg, D.~W., Lang, D., \& Goodman, J. 2013, PASP, 125, 306

\bibitem[{Fulton {et~al.}(2018)Fulton, Petigura, Blunt, \&
  Sinukoff}]{Fulton:2018gq}
Fulton, B.~J., Petigura, E.~A., Blunt, S., \& Sinukoff, E. 2018, PASP, 130,
  044504

\bibitem[{{Gaia Collaboration} {et~al.}(2018){Gaia Collaboration}, Brown,
  Vallenari, Prusti, de~Bruijne, Babusiaux, Bailer-Jones, Biermann, Evans,
  Eyer, Jansen, Jordi, Klioner, Lammers, Lindegren, Luri, Mignard, Panem,
  Pourbaix, Randich, Sartoretti, Siddiqui, Soubiran, van Leeuwen, Walton,
  Arenou, Bastian, Cropper, Drimmel, Katz, Lattanzi, Bakker, Cacciari,
  Casta{\~n}eda, Chaoul, Cheek, De~Angeli, Fabricius, Guerra, Holl, Masana,
  Messineo, Mowlavi, Nienartowicz, Panuzzo, Portell, Riello, Seabroke, Tanga,
  Th{\'e}venin, Gracia-Abril, Comoretto, Garcia-Reinaldos, Teyssier, Altmann,
  Andrae, Audard, Bellas-Velidis, Benson, Berthier, Blomme, Burgess, Busso,
  Carry, Cellino, Clementini, Clotet, Creevey, Davidson, De~Ridder, Delchambre,
  Dell{\textquoteright}Oro, Ducourant, Fern{\'a}ndez-Hern{\'a}ndez, Fouesneau,
  Fr{\'e}mat, Galluccio, Garc{\'\i}a-Torres, Gonz{\'a}lez-N{\'u}{\~n}ez,
  Gonz{\'a}lez-Vidal, Gosset, Guy, Halbwachs, Hambly, Harrison, Hern{\'a}ndez,
  Hestroffer, Hodgkin, Hutton, Jasniewicz, Jean-Antoine-Piccolo, Jordan, Korn,
  Krone-Martins, Lanzafame, Lebzelter, L{\"o}ffler, Manteiga, Marrese,
  Mart{\'\i}n-Fleitas, Moitinho, Mora, Muinonen, Osinde, Pancino, Pauwels,
  Petit, Recio-Blanco, Richards, Rimoldini, Robin, Sarro, Siopis, Smith,
  Sozzetti, S{\"u}veges, Torra, van Reeven, Abbas, Abreu~Aramburu, Accart,
  Aerts, Altavilla, {\'A}lvarez, Alvarez, Alves, Anderson, Andrei,
  Anglada~Varela, Antiche, Antoja, Arcay, Astraatmadja, Bach, Baker,
  Balaguer-N{\'u}{\~n}ez, Balm, Barache, Barata, Barbato, Barblan, Barklem,
  Barrado, Barros, Barstow, Bartholom{\'e}~Mu{\~n}oz, Bassilana, Becciani,
  Bellazzini, Berihuete, Bertone, Bianchi, Bienaym{\'e}, Blanco-Cuaresma, Boch,
  Boeche, Bombrun, Borrachero, Bossini, Bouquillon, Bourda, Bragaglia,
  Bramante, Breddels, Bressan, Brouillet, Br{\"u}semeister, Brugaletta,
  Bucciarelli, Burlacu, Busonero, Butkevich, Buzzi, Caffau, Cancelliere,
  Cannizzaro, Cantat-Gaudin, Carballo, Carlucci, Carrasco, Casamiquela,
  Castellani, Castro-Ginard, Charlot, Chemin, Chiavassa, Cocozza, Costigan,
  Cowell, Crifo, Crosta, Crowley, Cuypers, Dafonte, Damerdji, Dapergolas,
  David, David, de~Laverny, De~Luise, De~March, de~Martino, de~Souza,
  de~Torres, Debosscher, del Pozo, Delbo, Delgado, Delgado, Di~Matteo, Diakite,
  Diener, Distefano, Dolding, Drazinos, Dur{\'a}n, Edvardsson, Enke, Eriksson,
  Esquej, Eynard~Bontemps, Fabre, Fabrizio, Faigler, Falc{\~a}o,
  Farr{\`a}s~Casas, Federici, Fedorets, Fernique, Figueras, Filippi, Findeisen,
  Fonti, Fraile, Fraser, Fr{\'e}zouls, Gai, Galleti, Garabato,
  Garc{\'\i}a-Sedano, Garofalo, Garralda, Gavel, Gavras, Gerssen, Geyer,
  Giacobbe, Gilmore, Girona, Giuffrida, Glass, Gomes, Granvik, Gueguen,
  Guerrier, Guiraud, Guti{\'e}rrez-S{\'a}nchez, Haigron, Hatzidimitriou,
  Hauser, Haywood, Heiter, Helmi, Heu, Hilger, Hobbs, Hofmann, Holland, Huckle,
  Hypki, Icardi, Jan{\ss}en, Jevardat~de Fombelle, Jonker, Juh{\'a}sz, Julbe,
  Karampelas, Kewley, Klar, Kochoska, Kohley, Kolenberg, Kontizas, Kontizas,
  Koposov, \& Kordopatis}]{GaiaCollaboration:2018io}
{Gaia Collaboration}, Brown, A. G.~A., Vallenari, A., {et~al.} 2018, A{\&}A,
  616, A1

\bibitem[{Gandolfi {et~al.}(2018)Gandolfi, Barrag{\'a}n, Livingston, Fridlund,
  Justesen, Redfield, Fossati, Mathur, Grziwa, Cabrera, Garc{\'\i}a, Persson,
  Van~Eylen, Hatzes, Hidalgo, Albrecht, Bugnet, Cochran, Csizmadia, Deeg,
  Eigm{\"u}ller, Endl, Erikson, Esposito, Guenther, Korth, Luque,
  Monta{\~n}es~Rodr{\'\i}guez, Nespral, Nowak, P{\"a}tzold, \&
  Prieto-Arranz}]{Gandolfi:2018cg}
Gandolfi, D., Barrag{\'a}n, O., Livingston, J.~H., {et~al.} 2018, A{\&}A, 619,
  L10

\bibitem[{Garc{\'\i}a~Mu{\~n}oz {et~al.}(2020)Garc{\'\i}a~Mu{\~n}oz,
  Youngblood, Fossati, Gandolfi, Cabrera, \& Rauer}]{GarciaMunoz:2020dl}
Garc{\'\i}a~Mu{\~n}oz, A., Youngblood, A., Fossati, L., {et~al.} 2020,
  Astrophys. J., 888, L21

\bibitem[{Gray {et~al.}(2006)Gray, Corbally, Garrison, McFadden, Bubar,
  McGahee, O'Donoghue, \& Knox}]{Gray:2006ca}
Gray, R.~O., Corbally, C.~J., Garrison, R.~F., {et~al.} 2006, AJ, 132, 161

\bibitem[{{Hansen}(2017)}]{han17}
{Hansen}, B. M.~S. 2017, \mnras, 467, 1531

\bibitem[{{Hansen} \& {Murray}(2012)}]{han12}
{Hansen}, B. M.~S. \& {Murray}, N. 2012, \apj, 751, 158

\bibitem[{Heintz(1978)}]{Heintz:1978wp}
Heintz, W.~D. 1978, Geophysics and Astrophysics Monographs, 15

\bibitem[{{Herman} {et~al.}(2019){Herman}, {Zhu}, \& {Wu}}]{her19}
{Herman}, M.~K., {Zhu}, W., \& {Wu}, Y. 2019, \aj, 157, 248

\bibitem[{Huang {et~al.}(2018)Huang, Burt, Vanderburg, G{\"u}nther, Shporer,
  Dittmann, Winn, Wittenmyer, Sha, Kane, Ricker, Vanderspek, Latham, Seager,
  Jenkins, Caldwell, Collins, Guerrero, Smith, Quinn, Udry, Pepe, Bouchy,
  S{\'e}gransan, Lovis, Ehrenreich, Marmier, Mayor, Wohler, Haworth, Morgan,
  Fausnaugh, Ciardi, Christiansen, Charbonneau, Dragomir, Deming, Glidden,
  Levine, McCullough, Yu, Narita, Nguyen, Morton, Pepper, P{\'a}l, Rodriguez,
  Stassun, Torres, Sozzetti, Doty, Christensen-Dalsgaard, Laughlin, Clampin,
  Bean, Buchhave, Bakos, Sato, Ida, Kaltenegger, Pall{\'e}, Sasselov, Butler,
  Lissauer, Ge, \& Rinehart}]{Huang:2018dg}
Huang, C.~X., Burt, J., Vanderburg, A., {et~al.} 2018, Astrophys. J., 868, L39

\bibitem[{{Huang} {et~al.}(2017){Huang}, {Petrovich}, \& {Deibert}}]{hua17}
{Huang}, C.~X., {Petrovich}, C., \& {Deibert}, E. 2017, \aj, 153, 210

\bibitem[{{Hut}(1981)}]{hut81}
{Hut}, P. 1981, \aap, 99, 126

\bibitem[{Jones {et~al.}(2002)Jones, Paul~Butler, Tinney, Marcy, Penny,
  McCarthy, Carter, \& Pourbaix}]{Jones:2002cq}
Jones, H. R.~A., Paul~Butler, R., Tinney, C.~G., {et~al.} 2002, MNRAS, 333, 871

\bibitem[{{Juri{\'c}} \& {Tremaine}(2008)}]{jur08}
{Juri{\'c}}, M. \& {Tremaine}, S. 2008, \apj, 686, 603

\bibitem[{Kervella {et~al.}(2019)Kervella, Arenou, Mignard, \&
  Th{\'e}venin}]{Kervella:2019bw}
Kervella, P., Arenou, F., Mignard, F., \& Th{\'e}venin, F. 2019, A{\&}A, 623,
  A72

\bibitem[{{Kozai}(1962)}]{koz62}
{Kozai}, Y. 1962, \aj, 67, 591

\bibitem[{{Lai} \& {Pu}(2017)}]{lai17}
{Lai}, D. \& {Pu}, B. 2017, \aj, 153, 42

\bibitem[{{Lee} {et~al.}(2014){Lee}, {Chiang}, \& {Ormel}}]{lee14}
{Lee}, E.~J., {Chiang}, E., \& {Ormel}, C.~W. 2014, \apj, 797, 95

\bibitem[{{Li} {et~al.}(2014){Li}, {Naoz}, {Holman}, \& {Loeb}}]{li14}
{Li}, G., {Naoz}, S., {Holman}, M., \& {Loeb}, A. 2014, \apj, 791, 86

\bibitem[{{Lidov}(1962)}]{lid62}
{Lidov}, M.~L. 1962, \planss, 9, 719

\bibitem[{Lindegren(2019)}]{Lindegren:2019iv}
Lindegren, L. 2019, A{\&}A, 633, A1

\bibitem[{Lindegren(2020)}]{Lindegren:2020ik}
Lindegren, L. 2020, A{\&}A, 637, C5

\bibitem[{Lindegren {et~al.}(2018)Lindegren, Hern{\'a}ndez, Bombrun, Klioner,
  Bastian, Ramos-Lerate, de~Torres, Steidelm{\"u}ller, Stephenson, Hobbs,
  Lammers, Biermann, Geyer, Hilger, Michalik, Stampa, McMillan, Casta{\~n}eda,
  Clotet, Comoretto, Davidson, Fabricius, Gracia, Hambly, Hutton, Mora,
  Portell, van Leeuwen, Abbas, Abreu, Altmann, Andrei, Anglada,
  Balaguer-N{\'u}{\~n}ez, Barache, Becciani, Bertone, Bianchi, Bouquillon,
  Bourda, Br{\"u}semeister, Bucciarelli, Busonero, Buzzi, Cancelliere,
  Carlucci, Charlot, Cheek, Crosta, Crowley, de~Bruijne, de~Felice, Drimmel,
  Esquej, Fienga, Fraile, Gai, Garralda, Gonz{\'a}lez-Vidal, Guerra, Hauser,
  Hofmann, Holl, Jordan, Lattanzi, Lenhardt, Liao, Licata, Lister, L{\"o}ffler,
  Marchant, Mart{\'\i}n-Fleitas, Messineo, Mignard, Morbidelli, Poggio, Riva,
  Rowell, Salguero, Sarasso, Sciacca, Siddiqui, Smart, Spagna, Steele, Taris,
  Torra, van Elteren, van Reeven, \& Vecchiato}]{Lindegren:2018gy}
Lindegren, L., Hern{\'a}ndez, J., Bombrun, A., {et~al.} 2018, A{\&}A, 616, A2

\bibitem[{{Lithwick} \& {Wu}(2011)}]{lit11}
{Lithwick}, Y. \& {Wu}, Y. 2011, \apj, 739, 31

\bibitem[{Lo~Curto {et~al.}(2015)Lo~Curto, Pepe, Avila, Boffin, Bovay,
  Chazelas, Coffinet, Fleury, Hughes, Lovis, Maire, Manescau, Pasquini, Rihs,
  Sinclaire, \& Udry}]{LoCurto:2015vd}
Lo~Curto, G., Pepe, F., Avila, G., {et~al.} 2015, The Messenger, 162, 9

\bibitem[{{Maness} {et~al.}(2007){Maness}, {Marcy}, {Ford}, {Hauschildt},
  {Shreve}, {Basri}, {Butler}, \& {Vogt}}]{man07}
{Maness}, H.~L., {Marcy}, G.~W., {Ford}, E.~B., {et~al.} 2007, \pasp, 119, 90

\bibitem[{{Masuda} {et~al.}(2020){Masuda}, {Winn}, \& {Kawahara}}]{masu20}
{Masuda}, K., {Winn}, J.~N., \& {Kawahara}, H. 2020, \aj, 159, 38

\bibitem[{Mayor {et~al.}(2003)Mayor, Pepe, Queloz, Bouchy, Rupprecht, Lo~Curto,
  Avila, Benz, Bertaux, Bonfils, Dall, Dekker, Delabre, Eckert, Fleury,
  Gilliotte, Gojak, Guzman, Kohler, Lizon, Longinotti, Lovis, Megevand,
  Pasquini, Reyes, Sivan, Sosnowska, Soto, Udry, van Kesteren, Weber, \&
  Weilenmann}]{Mayor:2003wv}
Mayor, M., Pepe, F., Queloz, D., {et~al.} 2003, The Messenger, 114, 20

\bibitem[{McArthur {et~al.}(2010)McArthur, Benedict, Barnes, Martioli,
  Korzennik, Nelan, \& Butler}]{McArthur:2010jc}
McArthur, B.~E., Benedict, G.~F., Barnes, R., {et~al.} 2010, Astrophys. J.,
  715, 1203

\bibitem[{Mills \& Fabrycky(2017)}]{Mills:2017gw}
Mills, S.~M. \& Fabrycky, D.~C. 2017, AJ, 153, 45

\bibitem[{{Naoz}(2016)}]{nao16}
{Naoz}, S. 2016, \araa, 54, 441

\bibitem[{{Naoz} {et~al.}(2011){Naoz}, {Farr}, {Lithwick}, {Rasio}, \&
  {Teyssandier}}]{nao11}
{Naoz}, S., {Farr}, W.~M., {Lithwick}, Y., {Rasio}, F.~A., \& {Teyssandier}, J.
  2011, \nat, 473, 187

\bibitem[{{Nielsen} {et~al.}(2020){Nielsen}, {De Rosa}, {Wang}, {Sahlmann},
  {Kalas}, {Duch{\^e}ne}, {Rameau}, {Marley}, {Saumon}, {Macintosh},
  {Millar-Blanchaer}, {Nguyen}, {Ammons}, {Bailey}, {Barman}, {Bulger},
  {Chilcote}, {Cotten}, {Doyon}, {Esposito}, {Fitzgerald}, {Follette},
  {Gerard}, {Goodsell}, {Graham}, {Greenbaum}, {Hibon}, {Hung}, {Ingraham},
  {Konopacky}, {Larkin}, {Maire}, {Marchis}, {Marois}, {Metchev},
  {Oppenheimer}, {Palmer}, {Patience}, {Perrin}, {Poyneer}, {Pueyo}, {Rajan},
  {Rantakyr{\"o}}, {Ruffio}, {Savransky}, {Schneider}, {Sivaramakrishnan},
  {Song}, {Soummer}, {Thomas}, {Wallace}, {Ward-Duong}, {Wiktorowicz}, \&
  {Wolff}}]{nielsen:2020}
{Nielsen}, E.~L., {De Rosa}, R.~J., {Wang}, J.~J., {et~al.} 2020, \aj, 159, 71

\bibitem[{Noecker {et~al.}(2016)Noecker, Zhao, Demers, Trauger, Guyon, \&
  Jeremy~Kasdin}]{Noecker:2016hp}
Noecker, M.~C., Zhao, F., Demers, R., {et~al.} 2016, J. Astron. Telesc.
  Instrum. Syst., 2, 011001

\bibitem[{Ochsenbein {et~al.}(2000)Ochsenbein, Bauer, \&
  Marcout}]{Ochsenbein:2000fm}
Ochsenbein, F., Bauer, P., \& Marcout, J. 2000, A{\&}AS, 143, 23

\bibitem[{Pecaut \& Mamajek(2013)}]{Pecaut:2013ej}
Pecaut, M.~J. \& Mamajek, E.~E. 2013, ApJS, 208, 9

\bibitem[{{Petrovich}(2015{\natexlab{a}})}]{pet15}
{Petrovich}, C. 2015{\natexlab{a}}, \apj, 805, 75

\bibitem[{{Petrovich}(2015{\natexlab{b}})}]{petr15}
{Petrovich}, C. 2015{\natexlab{b}}, \apj, 808, 120

\bibitem[{{Petrovich} \& {Tremaine}(2016)}]{pet16}
{Petrovich}, C. \& {Tremaine}, S. 2016, \apj, 829, 132

\bibitem[{Price-Whelan {et~al.}(2017)Price-Whelan, Hogg, Foreman-Mackey, \&
  Rix}]{PriceWhelan:2017br}
Price-Whelan, A.~M., Hogg, D.~W., Foreman-Mackey, D., \& Rix, H.-W. 2017, ApJ,
  837, 20

\bibitem[{{Rasio} \& {Ford}(1996)}]{ras96}
{Rasio}, F.~A. \& {Ford}, E.~B. 1996, Science, 274, 954

\bibitem[{Reffert \& Quirrenbach(2011)}]{Reffert:2011ca}
Reffert, S. \& Quirrenbach, A. 2011, A{\&}A, 527, A140

\bibitem[{Ricker {et~al.}(2015)Ricker, Winn, Vanderspek, Latham, Bakos, Bean,
  Berta-Thompson, Brown, Buchhave, Butler, Butler, Chaplin, Charbonneau,
  Christensen-Dalsgaard, Clampin, Deming, Doty, De~Lee, Dressing, Dunham, Endl,
  Fressin, Ge, Henning, Holman, Howard, Ida, Jenkins, Jernigan, Johnson,
  Kaltenegger, Kawai, Kjeldsen, Laughlin, Levine, Lin, Lissauer, MacQueen,
  Marcy, McCullough, Morton, Narita, Paegert, Pall{\'e}, Pepe, Pepper,
  Quirrenbach, Rinehart, Sasselov, Sato, Seager, Sozzetti, Stassun, Sullivan,
  Szentgyorgyi, Torres, Udry, \& Villasenor}]{Ricker:2015ie}
Ricker, G.~R., Winn, J.~N., Vanderspek, R., {et~al.} 2015, JATIS, 1, 014003

\bibitem[{{Rogers} \& {Seager}(2010)}]{rog10}
{Rogers}, L.~A. \& {Seager}, S. 2010, \apj, 716, 1208

\bibitem[{Sahlmann {et~al.}(2011)Sahlmann, Segransan, Queloz, Udry, Santos,
  Marmier, Mayor, Naef, Pepe, \& Zucker}]{Sahlmann:2010hh}
Sahlmann, J., Segransan, D., Queloz, D., {et~al.} 2011, A{\&}A, 525, A95

\bibitem[{{Teyssandier} {et~al.}(2013){Teyssandier}, {Naoz}, {Lizarraga}, \&
  {Rasio}}]{tey13}
{Teyssandier}, J., {Naoz}, S., {Lizarraga}, I., \& {Rasio}, F.~A. 2013, \apj,
  779, 166

\bibitem[{van Leeuwen(2007{\natexlab{a}})}]{vanLeeuwen:2007du}
van Leeuwen, F. 2007{\natexlab{a}}, Astrophysics and Space Science Library, 350

\bibitem[{van Leeuwen(2007{\natexlab{b}})}]{vanLeeuwen:2007dc}
van Leeuwen, F. 2007{\natexlab{b}}, A{\&}A, 474, 653

\bibitem[{Virtanen {et~al.}(2020)Virtanen, Gommers, Oliphant, Haberland, Reddy,
  Cournapeau, Burovski, Peterson, Weckesser, Bright, van~der Walt, Brett,
  Wilson, Millman, Mayorov, Nelson, Jones, Kern, Larson, Carey, Polat, Feng,
  Moore, VanderPlas, Laxalde, Perktold, Cimrman, Henriksen, Quintero, Harris,
  Archibald, Ribeiro, Pedregosa, van Mulbregt, \&
  Contributors}]{Virtanen:2020cp}
Virtanen, P., Gommers, R., Oliphant, T.~E., {et~al.} 2020, Nature Methods, 17,
  261

\bibitem[{{Walsh} {et~al.}(2011){Walsh}, {Morbidelli}, {Raymond}, {O'Brien}, \&
  {Mandell}}]{wal11}
{Walsh}, K.~J., {Morbidelli}, A., {Raymond}, S.~N., {O'Brien}, D.~P., \&
  {Mandell}, A.~M. 2011, \nat, 475, 206

\bibitem[{Wenger {et~al.}(2000)Wenger, Ochsenbein, Egret, Dubois, Bonnarel,
  Borde, Genova, Jasniewicz, Lalo{\"e}, Lesteven, \& Monier}]{Wenger:2000ef}
Wenger, M., Ochsenbein, F., Egret, D., {et~al.} 2000, A{\&}AS, 143, 9

\bibitem[{{Wu} \& {Lithwick}(2011)}]{wu11}
{Wu}, Y. \& {Lithwick}, Y. 2011, \apj, 735, 109

\bibitem[{{Wu} \& {Murray}(2003)}]{wu03}
{Wu}, Y. \& {Murray}, N. 2003, \apj, 589, 605

\bibitem[{{Xuan} \& {Wyatt}(2020)}]{Xuan:2020aa}
{Xuan}, J.~W. \& {Wyatt}, M.~C. 2020, arXiv e-prints, arXiv:2007.01871

\bibitem[{{Yokoyama} {et~al.}(2003){Yokoyama}, {Santos}, {Cardin}, \&
  {Winter}}]{yoko03}
{Yokoyama}, T., {Santos}, M.~T., {Cardin}, G., \& {Winter}, O.~C. 2003, \aap,
  401, 763

\bibitem[{Zhu \& Wu(2018)}]{Zhu:2018js}
Zhu, W. \& Wu, Y. 2018, AJ, 156, 92

\bibitem[{Zurlo {et~al.}(2018)Zurlo, Mesa, Desidera, Messina, Gratton, Moutou,
  Beuzit, Biller, Boccaletti, Bonavita, Bonnefoy, Bhowmik, Brandner, Buenzli,
  Chauvin, Cudel, D'Orazi, Feldt, Hagelberg, Janson, Lagrange, Langlois,
  Lannier, Lavie, Lazzoni, Maire, Meyer, Mouillet, Peretti, Perrot, Potiron,
  Salter, Schmidt, Sissa, Vigan, Delboulbe, Petit, Ramos, Rigal, \&
  Rochat}]{Zurlo:2018da}
Zurlo, A., Mesa, D., Desidera, S., {et~al.} 2018, MNRAS, 480, 35

\end{thebibliography}

\longtab{
\begin{longtable}{cccc}
\caption{Radial Velocities of $\pi$ Mensae \label{tbl:rvs}}\\
\hline\hline
BJD$_{\rm TDB}$ & RV & $\pm\sigma$ & Instrument \\
(BJD-2450000)& km\,s$^{-1}$ & km\,s$^{-1}$ & \\
\hline
\endfirsthead
\caption{continued.}\\
\hline\hline
BJD$_{\rm TDB}$ & RV & $\pm\sigma$ & Instrument\\
(BJD-2450000)& km\,s$^{-1}$ & km\,s$^{-1}$ & \\
\hline
\endhead
\hline
\endfoot
829.9937 & $-0.01292$ & 0.00219 & UCLES\\
1119.2511 & $-0.04255$ & 0.00460 & UCLES\\
1236.0336 & $-0.05135$ & 0.00271 & UCLES\\
1411.3257 & $-0.05494$ & 0.00283 & UCLES\\
1473.2677 & $-0.05234$ & 0.00217 & UCLES\\
1526.0812 & $-0.06413$ & 0.00218 & UCLES\\
1527.0828 & $-0.06061$ & 0.00202 & UCLES\\
1530.1287 & $-0.05895$ & 0.00222 & UCLES\\
1629.9124 & $-0.06345$ & 0.00252 & UCLES\\
1683.8430 & $-0.06961$ & 0.00230 & UCLES\\
1828.1883 & $-0.03746$ & 0.00211 & UCLES\\
1919.0997 & $-0.01037$ & 0.00332 & UCLES\\
1921.1391 & $-0.01360$ & 0.00225 & UCLES\\
1983.9198 & $0.02081$ & 0.00241 & UCLES\\
2060.8404 & $0.16567$ & 0.00214 & UCLES\\
2092.3374 & $0.23967$ & 0.00217 & UCLES\\
2093.3522 & $0.24012$ & 0.00205 & UCLES\\
2127.3286 & $0.31441$ & 0.00269 & UCLES\\
2128.3364 & $0.31471$ & 0.00198 & UCLES\\
2130.3390 & $0.31686$ & 0.00314 & UCLES\\
2151.2924 & $0.33608$ & 0.00220 & UCLES\\
2154.3050 & $0.32535$ & 0.00500 & UCLES\\
2187.1966 & $0.31137$ & 0.00191 & UCLES\\
2188.2366 & $0.31145$ & 0.00197 & UCLES\\
2189.2226 & $0.30778$ & 0.00172 & UCLES\\
2190.1455 & $0.30929$ & 0.00187 & UCLES\\
2387.8714 & $0.12716$ & 0.00162 & UCLES\\
2510.3074 & $0.06962$ & 0.00211 & UCLES\\
2592.1264 & $0.04620$ & 0.00161 & UCLES\\
2599.1554 & $0.04655$ & 0.00580 & UCLES\\
2654.0993 & $0.04430$ & 0.00226 & UCLES\\
2708.9851 & $0.03204$ & 0.01137 & UCLES\\
2751.9185 & $0.01617$ & 0.00206 & UCLES\\
2944.2248 & $-0.01499$ & 0.00196 & UCLES\\
3001.8304 & $10.65999$ & 0.00144 & HARPS1\\
3004.0755 & $-0.00633$ & 0.00193 & UCLES\\
3034.6073 & $10.66647$ & 0.00078 & HARPS1\\
3042.0787 & $-0.01145$ & 0.00206 & UCLES\\
3043.0181 & $-0.01550$ & 0.00223 & UCLES\\
3047.0501 & $-0.01096$ & 0.00216 & UCLES\\
3048.0985 & $-0.02070$ & 0.00173 & UCLES\\
3245.3116 & $-0.04242$ & 0.00250 & UCLES\\
3289.8697 & $10.64482$ & 0.00125 & HARPS1\\
3289.8708 & $10.64277$ & 0.00109 & HARPS1\\
3289.8718 & $10.64465$ & 0.00107 & HARPS1\\
3289.8729 & $10.64494$ & 0.00113 & HARPS1\\
3289.8739 & $10.64522$ & 0.00115 & HARPS1\\
3289.8750 & $10.64648$ & 0.00115 & HARPS1\\
3289.8760 & $10.64591$ & 0.00110 & HARPS1\\
3289.8771 & $10.64715$ & 0.00109 & HARPS1\\
3289.8782 & $10.64492$ & 0.00118 & HARPS1\\
3289.8792 & $10.64727$ & 0.00114 & HARPS1\\
3291.8721 & $10.63853$ & 0.00155 & HARPS1\\
3291.8731 & $10.64122$ & 0.00158 & HARPS1\\
3291.8742 & $10.63669$ & 0.00179 & HARPS1\\
3291.8752 & $10.64222$ & 0.00164 & HARPS1\\
3291.8763 & $10.63837$ & 0.00144 & HARPS1\\
3291.8773 & $10.63781$ & 0.00145 & HARPS1\\
3291.8784 & $10.63809$ & 0.00160 & HARPS1\\
3291.8794 & $10.63992$ & 0.00144 & HARPS1\\
3292.8677 & $10.64073$ & 0.00155 & HARPS1\\
3292.8688 & $10.63609$ & 0.00160 & HARPS1\\
3292.8698 & $10.63871$ & 0.00154 & HARPS1\\
3292.8709 & $10.63941$ & 0.00159 & HARPS1\\
3292.8719 & $10.63633$ & 0.00162 & HARPS1\\
3292.8730 & $10.63640$ & 0.00153 & HARPS1\\
3292.8740 & $10.64083$ & 0.00158 & HARPS1\\
3292.8751 & $10.63954$ & 0.00170 & HARPS1\\
3294.8677 & $10.63729$ & 0.00162 & HARPS1\\
3294.8688 & $10.63078$ & 0.00193 & HARPS1\\
3294.8699 & $10.63067$ & 0.00226 & HARPS1\\
3294.8709 & $10.63496$ & 0.00202 & HARPS1\\
3294.8719 & $10.63389$ & 0.00197 & HARPS1\\
3294.8730 & $10.63865$ & 0.00189 & HARPS1\\
3294.8741 & $10.63448$ & 0.00182 & HARPS1\\
3294.8751 & $10.63472$ & 0.00196 & HARPS1\\
3296.8641 & $10.63774$ & 0.00098 & HARPS1\\
3296.8651 & $10.63819$ & 0.00089 & HARPS1\\
3296.8662 & $10.63589$ & 0.00092 & HARPS1\\
3296.8673 & $10.63512$ & 0.00092 & HARPS1\\
3296.8683 & $10.63857$ & 0.00094 & HARPS1\\
3296.8694 & $10.63932$ & 0.00092 & HARPS1\\
3296.8704 & $10.63822$ & 0.00089 & HARPS1\\
3296.8715 & $10.63544$ & 0.00099 & HARPS1\\
3296.8726 & $10.63641$ & 0.00089 & HARPS1\\
3296.8736 & $10.63887$ & 0.00095 & HARPS1\\
3402.0353 & $-0.03540$ & 0.00092 & UCLES\\
3669.2440 & $-0.05787$ & 0.00097 & UCLES\\
4012.2497 & $-0.00468$ & 0.00095 & UCLES\\
4039.1697 & $0.00351$ & 0.00112 & UCLES\\
4047.7963 & $10.68929$ & 0.00048 & HARPS1\\
4049.7768 & $10.68624$ & 0.00051 & HARPS1\\
4051.7893 & $10.68977$ & 0.00060 & HARPS1\\
4051.7939 & $10.68940$ & 0.00056 & HARPS1\\
4053.8538 & $10.69325$ & 0.00030 & HARPS1\\
4053.8575 & $10.69180$ & 0.00039 & HARPS1\\
4055.8425 & $10.69515$ & 0.00040 & HARPS1\\
4055.8473 & $10.69406$ & 0.00039 & HARPS1\\
4224.8540 & $0.32680$ & 0.00104 & UCLES\\
4336.3154 & $0.23368$ & 0.00183 & UCLES\\
4337.2922 & $0.23800$ & 0.00157 & UCLES\\
4372.2709 & $0.19521$ & 0.00183 & UCLES\\
4425.2248 & $0.15771$ & 0.00136 & UCLES\\
4545.9432 & $0.09094$ & 0.00104 & UCLES\\
4841.0648 & $0.02104$ & 0.00147 & UCLES\\
4901.9414 & $0.00710$ & 0.00204 & UCLES\\
4905.9923 & $0.00450$ & 0.00160 & UCLES\\
4906.9749 & $0.00529$ & 0.00128 & UCLES\\
5106.2401 & $-0.01748$ & 0.00197 & UCLES\\
5111.8704 & $10.65706$ & 0.00035 & HARPS1\\
5111.8759 & $10.65740$ & 0.00032 & HARPS1\\
5115.8554 & $10.65853$ & 0.00053 & HARPS1\\
5115.8610 & $10.65833$ & 0.00056 & HARPS1\\
5123.8644 & $10.65530$ & 0.00050 & HARPS1\\
5123.8685 & $10.65491$ & 0.00053 & HARPS1\\
5123.8724 & $10.65498$ & 0.00051 & HARPS1\\
5124.8595 & $10.65930$ & 0.00045 & HARPS1\\
5124.8627 & $10.65859$ & 0.00046 & HARPS1\\
5124.8662 & $10.65778$ & 0.00050 & HARPS1\\
5126.8301 & $10.65737$ & 0.00038 & HARPS1\\
5126.8339 & $10.65731$ & 0.00037 & HARPS1\\
5126.8378 & $10.65875$ & 0.00036 & HARPS1\\
5128.8596 & $10.65778$ & 0.00036 & HARPS1\\
5128.8635 & $10.65980$ & 0.00039 & HARPS1\\
5128.8673 & $10.65939$ & 0.00040 & HARPS1\\
5129.8586 & $10.65742$ & 0.00043 & HARPS1\\
5129.8619 & $10.65821$ & 0.00045 & HARPS1\\
5129.8651 & $10.65683$ & 0.00051 & HARPS1\\
5133.8356 & $10.65746$ & 0.00055 & HARPS1\\
5133.8414 & $10.65766$ & 0.00053 & HARPS1\\
5135.8146 & $10.65615$ & 0.00028 & HARPS1\\
5135.8201 & $10.65676$ & 0.00029 & HARPS1\\
5138.8325 & $10.65905$ & 0.00035 & HARPS1\\
5138.8378 & $10.65965$ & 0.00033 & HARPS1\\
5140.8049 & $10.65861$ & 0.00040 & HARPS1\\
5140.8104 & $10.65888$ & 0.00036 & HARPS1\\
5164.6598 & $10.65775$ & 0.00054 & HARPS1\\
5164.6631 & $10.65789$ & 0.00051 & HARPS1\\
5164.6664 & $10.65849$ & 0.00049 & HARPS1\\
5166.6152 & $10.65722$ & 0.00036 & HARPS1\\
5166.6185 & $10.65549$ & 0.00037 & HARPS1\\
5166.6218 & $10.65551$ & 0.00038 & HARPS1\\
5167.6346 & $10.65586$ & 0.00046 & HARPS1\\
5167.6372 & $10.65488$ & 0.00045 & HARPS1\\
5167.6399 & $10.65666$ & 0.00047 & HARPS1\\
5167.6426 & $10.65682$ & 0.00047 & HARPS1\\
5168.6583 & $10.65565$ & 0.00039 & HARPS1\\
5168.6610 & $10.65445$ & 0.00038 & HARPS1\\
5168.6637 & $10.65541$ & 0.00038 & HARPS1\\
5168.6664 & $10.65462$ & 0.00039 & HARPS1\\
5169.6633 & $10.65803$ & 0.00047 & HARPS1\\
5169.6659 & $10.65657$ & 0.00047 & HARPS1\\
5169.6686 & $10.65827$ & 0.00049 & HARPS1\\
5169.6713 & $10.65756$ & 0.00046 & HARPS1\\
5170.2379 & $-0.02015$ & 0.00134 & UCLES\\
5202.0552 & $-0.01497$ & 0.00192 & UCLES\\
5252.9700 & $-0.03226$ & 0.00132 & UCLES\\
5279.4871 & $10.64590$ & 0.00049 & HARPS1\\
5279.4912 & $10.64588$ & 0.00055 & HARPS1\\
5279.4955 & $10.64490$ & 0.00050 & HARPS1\\
5521.2013 & $-0.04540$ & 0.00159 & UCLES\\
5585.9968 & $-0.03564$ & 0.00170 & UCLES\\
5664.8584 & $-0.04455$ & 0.00148 & UCLES\\
5846.2520 & $-0.04801$ & 0.00152 & UCLES\\
5898.1102 & $-0.04673$ & 0.00128 & UCLES\\
5899.1379 & $-0.04953$ & 0.00148 & UCLES\\
5962.0046 & $-0.03537$ & 0.00127 & UCLES\\
5965.0396 & $-0.03758$ & 0.00137 & UCLES\\
5966.9802 & $-0.03659$ & 0.00225 & UCLES\\
6382.9076 & $0.29641$ & 0.00146 & UCLES\\
6558.2636 & $0.14524$ & 0.00132 & UCLES\\
6619.7332 & $10.78203$ & 0.00044 & HARPS1\\
6619.7365 & $10.77999$ & 0.00042 & HARPS1\\
6619.7397 & $10.78321$ & 0.00041 & HARPS1\\
6623.7165 & $10.77919$ & 0.00059 & HARPS1\\
6623.7201 & $10.78064$ & 0.00059 & HARPS1\\
6623.7238 & $10.77880$ & 0.00055 & HARPS1\\
6625.7589 & $10.78069$ & 0.00042 & HARPS1\\
6625.7645 & $10.78020$ & 0.00044 & HARPS1\\
6628.7485 & $10.77690$ & 0.00053 & HARPS1\\
6628.7543 & $10.77752$ & 0.00056 & HARPS1\\
6632.7204 & $10.77407$ & 0.00053 & HARPS1\\
6632.7257 & $10.77475$ & 0.00042 & HARPS1\\
6700.5207 & $10.75419$ & 0.00042 & HARPS1\\
6700.5263 & $10.75477$ & 0.00043 & HARPS1\\
6708.9621 & $0.07376$ & 0.00191 & UCLES\\
6721.5294 & $10.74248$ & 0.00042 & HARPS1\\
6721.5352 & $10.74286$ & 0.00045 & HARPS1\\
6723.5057 & $10.74270$ & 0.00045 & HARPS1\\
6723.5113 & $10.74348$ & 0.00047 & HARPS1\\
6725.5061 & $10.74632$ & 0.00047 & HARPS1\\
6725.5114 & $10.74575$ & 0.00078 & HARPS1\\
6769.8676 & $0.05966$ & 0.00215 & UCLES\\
6770.8867 & $0.07212$ & 0.00201 & UCLES\\
6938.2419 & $0.02823$ & 0.00269 & UCLES\\
6969.1694 & $0.01900$ & 0.00169 & UCLES\\
6990.7343 & $10.68744$ & 0.00034 & HARPS1\\
6990.7399 & $10.68792$ & 0.00036 & HARPS1\\
6994.7825 & $10.68824$ & 0.00046 & HARPS1\\
6994.7881 & $10.68926$ & 0.00046 & HARPS1\\
6998.6820 & $10.68622$ & 0.00030 & HARPS1\\
7002.7817 & $10.68814$ & 0.00031 & HARPS1\\
7033.6369 & $10.68479$ & 0.00026 & HARPS1\\
7054.9742 & $0.01672$ & 0.00195 & UCLES\\
7094.9653 & $0.00757$ & 0.00162 & UCLES\\
7298.8532 & $10.67498$ & 0.00055 & HARPS2\\
7298.8582 & $10.67466$ & 0.00042 & HARPS2\\
7327.7558 & $10.67442$ & 0.00034 & HARPS2\\
7349.2184 & $-0.02386$ & 0.00194 & UCLES\\
7354.7837 & $10.66740$ & 0.00020 & HARPS2\\
7357.7259 & $10.67274$ & 0.00020 & HARPS2\\
7372.7051 & $10.66638$ & 0.00039 & HARPS2\\
7372.7090 & $10.66619$ & 0.00043 & HARPS2\\
7372.7128 & $10.66536$ & 0.00033 & HARPS2\\
7423.5918 & $10.66299$ & 0.00050 & HARPS2\\
7423.5979 & $10.66281$ & 0.00050 & HARPS2\\
7424.5866 & $10.66447$ & 0.00036 & HARPS2\\
7424.5924 & $10.66430$ & 0.00037 & HARPS2\\
7462.5179 & $10.66123$ & 0.00032 & HARPS2\\
7462.5235 & $10.66120$ & 0.00031 & HARPS2\\
7464.4999 & $10.66162$ & 0.00050 & HARPS2\\
7464.5038 & $10.66271$ & 0.00039 & HARPS2\\
7464.5075 & $10.66112$ & 0.00038 & HARPS2\\
8383.8964 & $10.99449$ & 0.00043 & HARPS2\\
8383.8995 & $10.99439$ & 0.00044 & HARPS2\\
8384.8140 & $10.99870$ & 0.00041 & HARPS2\\
8384.8171 & $11.00013$ & 0.00043 & HARPS2\\
8385.8134 & $11.00313$ & 0.00037 & HARPS2\\
8385.8165 & $11.00331$ & 0.00037 & HARPS2\\
8385.8985 & $11.00357$ & 0.00034 & HARPS2\\
8385.9017 & $11.00391$ & 0.00034 & HARPS2\\
8386.8961 & $11.00388$ & 0.00038 & HARPS2\\
8386.8993 & $11.00651$ & 0.00037 & HARPS2\\
8393.7901 & $11.01202$ & 0.00045 & HARPS2\\
8393.7933 & $11.01291$ & 0.00048 & HARPS2\\
8393.7965 & $11.01177$ & 0.00051 & HARPS2\\
8396.8747 & $11.01703$ & 0.00061 & HARPS2\\
8396.8778 & $11.01900$ & 0.00077 & HARPS2\\
8396.8811 & $11.01682$ & 0.00071 & HARPS2\\
8397.8870 & $11.01882$ & 0.00056 & HARPS2\\
8397.8902 & $11.01958$ & 0.00054 & HARPS2\\
8397.8933 & $11.02060$ & 0.00053 & HARPS2\\
8398.8927 & $11.02130$ & 0.00058 & HARPS2\\
8398.8955 & $11.02131$ & 0.00063 & HARPS2\\
8398.8988 & $11.02145$ & 0.00087 & HARPS2\\
8405.8822 & $11.02551$ & 0.00137 & HARPS2\\
8405.8853 & $11.02839$ & 0.00132 & HARPS2\\
8405.8884 & $11.02791$ & 0.00121 & HARPS2\\
8407.8498 & $11.02717$ & 0.00058 & HARPS2\\
8407.8530 & $11.02821$ & 0.00060 & HARPS2\\
8407.8562 & $11.02881$ & 0.00067 & HARPS2\\
8415.7025 & $11.03223$ & 0.00050 & HARPS2\\
8415.7061 & $11.03368$ & 0.00049 & HARPS2\\
8415.7101 & $11.03352$ & 0.00052 & HARPS2\\
8415.7139 & $11.03194$ & 0.00046 & HARPS2\\
8415.7177 & $11.03310$ & 0.00045 & HARPS2\\
8415.7215 & $11.03276$ & 0.00046 & HARPS2\\
8415.7735 & $11.03186$ & 0.00045 & HARPS2\\
8415.7774 & $11.03311$ & 0.00047 & HARPS2\\
8415.7814 & $11.03254$ & 0.00045 & HARPS2\\
8415.7850 & $11.03203$ & 0.00042 & HARPS2\\
8415.7889 & $11.03144$ & 0.00041 & HARPS2\\
8415.7927 & $11.03213$ & 0.00038 & HARPS2\\
8416.7930 & $11.03509$ & 0.00098 & HARPS2\\
8416.7967 & $11.03324$ & 0.00123 & HARPS2\\
8416.8006 & $11.03463$ & 0.00092 & HARPS2\\
8424.8539 & $11.03525$ & 0.00051 & HARPS2\\
8424.8577 & $11.03490$ & 0.00048 & HARPS2\\
8424.8614 & $11.03568$ & 0.00051 & HARPS2\\
8424.8653 & $11.03419$ & 0.00054 & HARPS2\\
8425.8083 & $11.03196$ & 0.00037 & HARPS2\\
8425.8121 & $11.03160$ & 0.00039 & HARPS2\\
8425.8160 & $11.03218$ & 0.00039 & HARPS2\\
8425.8438 & $11.03257$ & 0.00038 & HARPS2\\
8425.8477 & $11.03162$ & 0.00037 & HARPS2\\
8425.8514 & $11.03205$ & 0.00038 & HARPS2\\
8425.8554 & $11.03153$ & 0.00038 & HARPS2\\
8425.8591 & $11.03323$ & 0.00039 & HARPS2\\
8425.8630 & $11.03209$ & 0.00038 & HARPS2\\
8426.7951 & $11.03258$ & 0.00037 & HARPS2\\
8426.7989 & $11.03219$ & 0.00036 & HARPS2\\
8426.8028 & $11.03138$ & 0.00036 & HARPS2\\
8426.8533 & $11.03105$ & 0.00049 & HARPS2\\
8426.8569 & $11.03062$ & 0.00045 & HARPS2\\
8426.8612 & $11.02976$ & 0.00042 & HARPS2\\
8427.7921 & $11.03100$ & 0.00037 & HARPS2\\
8427.7959 & $11.03033$ & 0.00038 & HARPS2\\
8427.7998 & $11.03143$ & 0.00038 & HARPS2\\
8427.8305 & $11.03121$ & 0.00034 & HARPS2\\
8427.8343 & $11.02977$ & 0.00033 & HARPS2\\
8427.8382 & $11.03073$ & 0.00034 & HARPS2\\
8428.8147 & $11.03211$ & 0.00038 & HARPS2\\
8428.8184 & $11.03260$ & 0.00037 & HARPS2\\
8428.8223 & $11.03253$ & 0.00037 & HARPS2\\
8428.8399 & $11.03265$ & 0.00036 & HARPS2\\
8428.8437 & $11.03298$ & 0.00034 & HARPS2\\
8428.8475 & $11.03274$ & 0.00036 & HARPS2\\
8428.8519 & $11.03384$ & 0.00037 & HARPS2\\
8428.8557 & $11.03299$ & 0.00038 & HARPS2\\
8428.8596 & $11.03318$ & 0.00036 & HARPS2\\
8436.7996 & $11.03314$ & 0.00066 & HARPS2\\
8436.8034 & $11.03298$ & 0.00061 & HARPS2\\
8436.8071 & $11.03258$ & 0.00061 & HARPS2\\
8438.8040 & $11.02768$ & 0.00043 & HARPS2\\
8438.8078 & $11.02671$ & 0.00040 & HARPS2\\
8438.8116 & $11.02615$ & 0.00040 & HARPS2\\
8442.7713 & $11.02614$ & 0.00036 & HARPS2\\
8442.7752 & $11.02578$ & 0.00036 & HARPS2\\
8442.7790 & $11.02620$ & 0.00037 & HARPS2\\
8447.7514 & $11.02096$ & 0.00031 & HARPS2\\
8447.7553 & $11.02125$ & 0.00031 & HARPS2\\
8447.7591 & $11.02054$ & 0.00032 & HARPS2\\
8448.8555 & $11.02063$ & 0.00039 & HARPS2\\
8448.8594 & $11.02085$ & 0.00040 & HARPS2\\
8448.8631 & $11.02109$ & 0.00040 & HARPS2\\
8451.7820 & $11.01131$ & 0.00042 & HARPS2\\
8451.7859 & $11.01200$ & 0.00044 & HARPS2\\
8451.7897 & $11.01214$ & 0.00046 & HARPS2\\
8455.7669 & $11.00983$ & 0.00043 & HARPS2\\
8455.7707 & $11.01003$ & 0.00041 & HARPS2\\
8455.7746 & $11.00980$ & 0.00043 & HARPS2\\
8467.7659 & $10.99464$ & 0.00042 & HARPS2\\
8467.7697 & $10.99328$ & 0.00040 & HARPS2\\
8467.7735 & $10.99477$ & 0.00040 & HARPS2\\
8473.7348 & $10.98596$ & 0.00033 & HARPS2\\
8473.7386 & $10.98594$ & 0.00032 & HARPS2\\
8473.7442 & $10.98604$ & 0.00034 & HARPS2\\
8473.7479 & $10.98547$ & 0.00034 & HARPS2\\
8473.7518 & $10.98588$ & 0.00033 & HARPS2\\
8484.7341 & $10.97020$ & 0.00038 & HARPS2\\
8484.7380 & $10.96975$ & 0.00039 & HARPS2\\
8486.6897 & $10.96800$ & 0.00054 & HARPS2\\
8486.6934 & $10.96832$ & 0.00052 & HARPS2\\
8487.6801 & $10.96859$ & 0.00034 & HARPS2\\
8487.6840 & $10.96873$ & 0.00035 & HARPS2\\
8488.6917 & $10.96610$ & 0.00044 & HARPS2\\
8488.6955 & $10.96719$ & 0.00045 & HARPS2\\
8489.7051 & $10.96437$ & 0.00061 & HARPS2\\
8489.7092 & $10.96520$ & 0.00061 & HARPS2\\
8490.7356 & $10.95853$ & 0.00057 & HARPS2\\
8490.7395 & $10.95907$ & 0.00053 & HARPS2\\
8491.6539 & $10.96268$ & 0.00041 & HARPS2\\
8491.7025 & $10.96197$ & 0.00046 & HARPS2\\
8508.5840 & $10.94137$ & 0.00052 & HARPS2\\
8508.5877 & $10.94177$ & 0.00051 & HARPS2\\
8508.5916 & $10.94164$ & 0.00050 & HARPS2\\
8535.5708 & $10.91178$ & 0.00041 & HARPS2\\
8535.5746 & $10.91129$ & 0.00044 & HARPS2\\
8591.4688 & $10.86539$ & 0.00053 & HARPS2\\
8591.4726 & $10.86581$ & 0.00053 & HARPS2\\
8591.4765 & $10.86589$ & 0.00051 & HARPS2\\
8626.4568 & $10.83914$ & 0.00070 & HARPS2\\
8626.4606 & $10.84032$ & 0.00068 & HARPS2\\
8626.4658 & $10.84027$ & 0.00074 & HARPS2\\
8626.4697 & $10.83888$ & 0.00068 & HARPS2\\
8767.8287 & $10.77613$ & 0.00044 & HARPS2\\
8767.8325 & $10.77577$ & 0.00044 & HARPS2\\
8767.8362 & $10.77534$ & 0.00041 & HARPS2\\
8856.7783 & $10.75115$ & 0.00072 & HARPS2\\
\end{longtable}
}

\longtab{
\begin{longtable}{cccccc}
\caption{{\it Hipparcos} astrometric measurements of the photocenter of the $\pi$ Mensae system \citep{vanLeeuwen:2007du}\label{tbl:iad}}\\
\hline\hline
$t-1991.25$ & $\Pi$ & $\cos\psi$ & $\sin\psi$ & $\delta\Lambda$ & $\sigma_\Lambda$ \\
yr & & & & mas & mas\\
\hline
\endfirsthead
\caption{continued.}\\
\hline\hline
$t-1991.25$ & $\Pi$ & $\cos\psi$ & $\sin\psi$ & $\delta\Lambda$ & $\sigma_\Lambda$ \\
yr & & & & mas & mas\\
\hline
\endhead
\hline
\endfoot
$-1.405$ & $0.659$ & $-0.0965$ & $-0.9953$ & $2.36$ & $1.28$\\
$-1.405$ & $0.659$ & $-0.0966$ & $-0.9953$ & $-2.71$ & $1.21$\\
$-1.405$ & $0.659$ & $-0.0960$ & $-0.9954$ & $-0.91$ & $1.18$\\
$-1.405$ & $0.659$ & $-0.0963$ & $-0.9954$ & $-0.66$ & $1.13$\\
$-1.260$ & $0.627$ & $-0.8763$ & $-0.4817$ & $0.23$ & $1.00$\\
$-1.260$ & $0.627$ & $-0.8759$ & $-0.4825$ & $0.87$ & $0.85$\\
$-1.260$ & $0.629$ & $-0.8748$ & $-0.4845$ & $0.56$ & $1.31$\\
$-1.260$ & $0.628$ & $-0.8753$ & $-0.4836$ & $-0.92$ & $0.92$\\
$-1.260$ & $0.626$ & $-0.8767$ & $-0.4810$ & $-0.40$ & $0.91$\\
$-1.260$ & $0.627$ & $-0.8758$ & $-0.4826$ & $-1.60$ & $1.17$\\
$-1.168$ & $-0.642$ & $-0.0193$ & $0.9998$ & $-1.41$ & $0.96$\\
$-1.168$ & $-0.642$ & $-0.0184$ & $0.9998$ & $-2.99$ & $1.81$\\
$-1.111$ & $0.660$ & $-0.8980$ & $0.4399$ & $-0.86$ & $1.55$\\
$-1.110$ & $0.665$ & $-0.8976$ & $0.4407$ & $-0.72$ & $1.35$\\
$-1.110$ & $0.666$ & $-0.8980$ & $0.4401$ & $-2.18$ & $1.24$\\
$-1.023$ & $-0.682$ & $0.7961$ & $0.6052$ & $2.47$ & $1.43$\\
$-1.023$ & $-0.682$ & $0.7960$ & $0.6054$ & $-0.33$ & $1.00$\\
$-1.023$ & $-0.684$ & $0.7974$ & $0.6034$ & $-1.17$ & $1.48$\\
$-0.949$ & $0.676$ & $-0.1226$ & $0.9925$ & $-0.16$ & $1.02$\\
$-0.949$ & $0.674$ & $-0.1203$ & $0.9927$ & $-0.58$ & $0.90$\\
$-0.949$ & $0.674$ & $-0.1196$ & $0.9928$ & $-1.86$ & $1.14$\\
$-0.949$ & $0.677$ & $-0.1241$ & $0.9923$ & $-1.09$ & $1.20$\\
$-0.949$ & $0.678$ & $-0.1258$ & $0.9921$ & $0.20$ & $1.08$\\
$-0.868$ & $-0.659$ & $0.9618$ & $-0.2738$ & $0.94$ & $1.13$\\
$-0.868$ & $-0.660$ & $0.9612$ & $-0.2758$ & $-1.22$ & $1.24$\\
$-0.868$ & $-0.658$ & $0.9620$ & $-0.2731$ & $-0.18$ & $1.32$\\
$-0.868$ & $-0.661$ & $0.9610$ & $-0.2766$ & $-3.01$ & $1.44$\\
$-0.780$ & $0.650$ & $0.8120$ & $0.5836$ & $-1.39$ & $0.90$\\
$-0.780$ & $0.645$ & $0.8162$ & $0.5778$ & $-0.61$ & $1.83$\\
$-0.780$ & $0.645$ & $0.8165$ & $0.5773$ & $0.29$ & $0.94$\\
$-0.780$ & $0.650$ & $0.8126$ & $0.5829$ & $1.52$ & $1.75$\\
$-0.780$ & $0.649$ & $0.8135$ & $0.5815$ & $-3.72$ & $1.71$\\
$-0.780$ & $0.648$ & $0.8140$ & $0.5808$ & $-0.27$ & $0.90$\\
$-0.701$ & $-0.653$ & $0.2569$ & $-0.9664$ & $0.84$ & $1.70$\\
$-0.701$ & $-0.655$ & $0.2543$ & $-0.9671$ & $0.01$ & $1.06$\\
$-0.701$ & $-0.658$ & $0.2506$ & $-0.9681$ & $-2.16$ & $1.62$\\
$-0.701$ & $-0.655$ & $0.2541$ & $-0.9672$ & $0.79$ & $1.63$\\
$-0.701$ & $-0.658$ & $0.2509$ & $-0.9680$ & $-2.69$ & $0.98$\\
$-0.532$ & $-0.686$ & $-0.7182$ & $-0.6958$ & $-0.24$ & $0.94$\\
$-0.532$ & $-0.684$ & $-0.7168$ & $-0.6973$ & $1.05$ & $1.03$\\
$-0.532$ & $-0.685$ & $-0.7178$ & $-0.6963$ & $-0.61$ & $1.25$\\
$-0.532$ & $-0.685$ & $-0.7175$ & $-0.6965$ & $-1.12$ & $1.27$\\
$-0.532$ & $-0.683$ & $-0.7158$ & $-0.6983$ & $1.24$ & $1.35$\\
$-0.532$ & $-0.682$ & $-0.7152$ & $-0.6989$ & $2.45$ & $1.14$\\
$-0.464$ & $0.675$ & $0.2777$ & $-0.9607$ & $0.26$ & $0.87$\\
$-0.464$ & $0.675$ & $0.2767$ & $-0.9609$ & $-2.51$ & $1.35$\\
$-0.463$ & $0.679$ & $0.2761$ & $-0.9611$ & $-0.10$ & $1.03$\\
$-0.463$ & $0.678$ & $0.2747$ & $-0.9615$ & $-1.10$ & $1.45$\\
$-0.374$ & $-0.654$ & $-0.9808$ & $0.1951$ & $-0.56$ & $1.28$\\
$-0.374$ & $-0.654$ & $-0.9808$ & $0.1950$ & $-2.08$ & $2.38$\\
$-0.373$ & $-0.646$ & $-0.9812$ & $0.1929$ & $-2.06$ & $1.54$\\
$-0.373$ & $-0.646$ & $-0.9811$ & $0.1933$ & $6.20$ & $2.47$\\
$-0.318$ & $0.633$ & $-0.6252$ & $-0.7805$ & $-0.68$ & $1.87$\\
$-0.318$ & $0.633$ & $-0.6255$ & $-0.7802$ & $-0.92$ & $0.97$\\
$-0.318$ & $0.633$ & $-0.6255$ & $-0.7803$ & $0.33$ & $1.79$\\
$-0.318$ & $0.633$ & $-0.6254$ & $-0.7803$ & $-1.85$ & $1.06$\\
$-0.226$ & $-0.633$ & $-0.3960$ & $0.9182$ & $-0.81$ & $1.11$\\
$-0.226$ & $-0.632$ & $-0.3973$ & $0.9177$ & $-0.04$ & $0.99$\\
$-0.226$ & $-0.631$ & $-0.3980$ & $0.9174$ & $0.15$ & $1.23$\\
$-0.226$ & $-0.633$ & $-0.3962$ & $0.9182$ & $-0.37$ & $0.90$\\
$-0.171$ & $0.642$ & $-0.9972$ & $0.0752$ & $-8.36$ & $1.45$\\
$-0.171$ & $0.643$ & $-0.9973$ & $0.0733$ & $0.33$ & $2.43$\\
$-0.083$ & $-0.674$ & $0.5237$ & $0.8519$ & $-2.91$ & $1.68$\\
$-0.081$ & $-0.670$ & $0.5247$ & $0.8513$ & $0.14$ & $0.89$\\
$-0.081$ & $-0.670$ & $0.5249$ & $0.8512$ & $-4.12$ & $1.72$\\
$-0.014$ & $0.679$ & $-0.5060$ & $0.8625$ & $2.48$ & $2.01$\\
$-0.014$ & $0.680$ & $-0.5067$ & $0.8621$ & $-0.66$ & $2.33$\\
$-0.014$ & $0.680$ & $-0.5071$ & $0.8619$ & $1.40$ & $1.71$\\
$-0.014$ & $0.681$ & $-0.5085$ & $0.8610$ & $0.29$ & $2.09$\\
$0.068$ & $-0.674$ & $0.9948$ & $0.1018$ & $2.12$ & $1.68$\\
$0.068$ & $-0.674$ & $0.9948$ & $0.1022$ & $0.79$ & $1.23$\\
$0.069$ & $-0.673$ & $0.9954$ & $0.0960$ & $-2.46$ & $4.20$\\
$0.152$ & $0.654$ & $0.4970$ & $0.8678$ & $2.63$ & $1.05$\\
$0.152$ & $0.651$ & $0.5006$ & $0.8657$ & $0.30$ & $1.53$\\
$0.152$ & $0.657$ & $0.4942$ & $0.8693$ & $-1.30$ & $2.38$\\
$0.152$ & $0.654$ & $0.4969$ & $0.8678$ & $1.52$ & $1.62$\\
$0.231$ & $-0.650$ & $0.6305$ & $-0.7762$ & $-2.21$ & $1.72$\\
$0.231$ & $-0.649$ & $0.6310$ & $-0.7757$ & $-0.49$ & $0.95$\\
$0.231$ & $-0.647$ & $0.6331$ & $-0.7741$ & $-1.64$ & $1.66$\\
$0.231$ & $-0.645$ & $0.6360$ & $-0.7717$ & $-0.22$ & $1.14$\\
$0.231$ & $-0.644$ & $0.6366$ & $-0.7712$ & $-1.38$ & $1.57$\\
$0.231$ & $-0.647$ & $0.6336$ & $-0.7737$ & $-2.60$ & $1.13$\\
$0.320$ & $0.663$ & $0.9999$ & $0.0102$ & $-2.51$ & $1.48$\\
$0.320$ & $0.663$ & $0.9999$ & $0.0103$ & $1.61$ & $1.20$\\
$0.320$ & $0.660$ & $1.0000$ & $0.0062$ & $-3.17$ & $1.49$\\
$0.320$ & $0.659$ & $1.0000$ & $0.0048$ & $-0.46$ & $1.14$\\
$0.401$ & $-0.676$ & $-0.3771$ & $-0.9262$ & $1.18$ & $2.47$\\
$0.401$ & $-0.680$ & $-0.3817$ & $-0.9243$ & $0.48$ & $2.75$\\
$0.401$ & $-0.678$ & $-0.3789$ & $-0.9254$ & $-0.25$ & $2.30$\\
$0.401$ & $-0.680$ & $-0.3816$ & $-0.9243$ & $-1.88$ & $1.67$\\
$0.401$ & $-0.676$ & $-0.3775$ & $-0.9260$ & $1.49$ & $2.61$\\
$0.401$ & $-0.677$ & $-0.3786$ & $-0.9256$ & $0.18$ & $1.99$\\
$0.476$ & $0.687$ & $0.6045$ & $-0.7966$ & $0.94$ & $1.57$\\
$0.476$ & $0.685$ & $0.6030$ & $-0.7977$ & $1.77$ & $2.43$\\
$0.476$ & $0.686$ & $0.6038$ & $-0.7971$ & $1.88$ & $1.93$\\
$0.476$ & $0.687$ & $0.6053$ & $-0.7960$ & $1.87$ & $1.42$\\
$0.564$ & $-0.673$ & $-0.9826$ & $-0.1859$ & $-2.03$ & $1.49$\\
$0.564$ & $-0.673$ & $-0.9825$ & $-0.1863$ & $0.52$ & $2.80$\\
$0.565$ & $-0.667$ & $-0.9825$ & $-0.1860$ & $-0.87$ & $1.06$\\
$0.625$ & $0.653$ & $-0.2838$ & $-0.9589$ & $2.52$ & $2.67$\\
$0.625$ & $0.653$ & $-0.2840$ & $-0.9588$ & $-2.32$ & $1.98$\\
$0.716$ & $-0.633$ & $-0.7166$ & $0.6974$ & $-0.26$ & $2.04$\\
$0.716$ & $-0.634$ & $-0.7155$ & $0.6986$ & $0.79$ & $2.49$\\
$0.716$ & $-0.633$ & $-0.7163$ & $0.6978$ & $-0.77$ & $1.46$\\
$0.716$ & $-0.634$ & $-0.7157$ & $0.6984$ & $-0.75$ & $2.19$\\
$0.769$ & $0.627$ & $-0.9528$ & $-0.3036$ & $-4.13$ & $2.25$\\
$0.769$ & $0.627$ & $-0.9525$ & $-0.3045$ & $-1.92$ & $2.74$\\
$0.769$ & $0.626$ & $-0.9529$ & $-0.3034$ & $0.20$ & $2.15$\\
$0.860$ & $-0.658$ & $0.1740$ & $0.9847$ & $-1.42$ & $1.59$\\
$0.860$ & $-0.656$ & $0.1724$ & $0.9850$ & $-0.17$ & $2.08$\\
$0.860$ & $-0.656$ & $0.1725$ & $0.9850$ & $-0.88$ & $1.40$\\
$0.860$ & $-0.656$ & $0.1726$ & $0.9850$ & $0.60$ & $1.10$\\
$0.860$ & $-0.657$ & $0.1740$ & $0.9848$ & $-0.63$ & $1.34$\\
$0.921$ & $0.666$ & $-0.7974$ & $0.6035$ & $0.37$ & $1.06$\\
$0.921$ & $0.666$ & $-0.7972$ & $0.6037$ & $2.79$ & $1.72$\\
$1.007$ & $-0.679$ & $0.8920$ & $0.4521$ & $1.29$ & $2.54$\\
$1.007$ & $-0.681$ & $0.8934$ & $0.4493$ & $0.50$ & $1.50$\\
$1.007$ & $-0.681$ & $0.8930$ & $0.4501$ & $-0.54$ & $1.94$\\
$1.007$ & $-0.678$ & $0.8913$ & $0.4533$ & $0.01$ & $1.61$\\
$1.007$ & $-0.681$ & $0.8930$ & $0.4501$ & $-6.56$ & $2.82$\\
$1.085$ & $0.670$ & $0.0865$ & $0.9963$ & $-2.52$ & $1.39$\\
$1.085$ & $0.670$ & $0.0865$ & $0.9963$ & $0.35$ & $2.85$\\
$1.085$ & $0.668$ & $0.0895$ & $0.9960$ & $-3.12$ & $1.46$\\
$1.085$ & $0.668$ & $0.0896$ & $0.9960$ & $0.75$ & $2.08$\\
$1.165$ & $-0.651$ & $0.8886$ & $-0.4587$ & $0.50$ & $2.50$\\
$1.165$ & $-0.650$ & $0.8892$ & $-0.4576$ & $0.08$ & $2.59$\\
$1.165$ & $-0.647$ & $0.8908$ & $-0.4545$ & $-2.54$ & $1.40$\\
$1.254$ & $0.646$ & $0.9193$ & $0.3935$ & $5.74$ & $2.60$\\
$1.254$ & $0.646$ & $0.9192$ & $0.3938$ & $3.69$ & $2.80$\\
$1.254$ & $0.652$ & $0.9160$ & $0.4011$ & $1.53$ & $1.41$\\
$1.254$ & $0.651$ & $0.9162$ & $0.4008$ & $1.18$ & $2.83$\\
$1.254$ & $0.648$ & $0.9179$ & $0.3969$ & $-0.83$ & $1.53$\\
$1.254$ & $0.648$ & $0.9179$ & $0.3968$ & $-1.81$ & $1.77$\\
$1.714$ & $0.627$ & $-0.7807$ & $-0.6249$ & $0.38$ & $1.21$\\
$1.714$ & $0.626$ & $-0.7815$ & $-0.6239$ & $2.70$ & $1.48$\\
$1.828$ & $0.643$ & $-0.9977$ & $0.0675$ & $1.70$ & $1.63$\\
$1.828$ & $0.644$ & $-0.9978$ & $0.0666$ & $1.61$ & $1.63$\\
\end{longtable}
}

\longtab{
\begin{longtable}{cccccc}
\caption{Predicted {\it Gaia} scan timings, orientations and parallax factors for the $\pi$ Mensae system.\label{tbl:gost}}\\
\hline\hline
$t-2015.5$ & MJD & $\Pi$ & $\sin\theta$ & $\cos\theta$ \\
yr & d & & & \\
\hline
\endfirsthead
\caption{continued.}\\
\hline\hline
$t-2015.5$ & MJD & $\Pi$ & $\sin\theta$ & $\cos\theta$ \\
yr & d & & & \\
\hline
\endhead
\hline
\endfoot
$-0.849$ & $56895.50$ & $0.71076$ & $0.90866$ & $-0.41754$\\
$-0.739$ & $56935.80$ & $0.71290$ & $0.46076$ & $-0.88753$\\
$-0.739$ & $56935.87$ & $0.71306$ & $0.45989$ & $-0.88798$\\
$-0.578$ & $56994.51$ & $0.66879$ & $-0.53471$ & $-0.84503$\\
$-0.477$ & $57031.22$ & $-0.66503$ & $-0.37964$ & $0.92513$\\
$-0.477$ & $57031.30$ & $-0.66370$ & $-0.38017$ & $0.92492$\\
$-0.417$ & $57053.47$ & $0.67270$ & $-0.99817$ & $0.06050$\\
$-0.416$ & $57053.54$ & $0.67386$ & $-0.99818$ & $0.06023$\\
$-0.319$ & $57089.18$ & $-0.70386$ & $0.61457$ & $0.78886$\\
$-0.318$ & $57089.26$ & $-0.70352$ & $0.61515$ & $0.78841$\\
$-0.151$ & $57150.41$ & $-0.69575$ & $0.99420$ & $-0.10758$\\
$-0.056$ & $57184.89$ & $0.68591$ & $0.67338$ & $0.73930$\\
$0.032$ & $57217.13$ & $-0.68494$ & $0.33270$ & $-0.94303$\\
$0.032$ & $57217.21$ & $-0.68608$ & $0.33006$ & $-0.94396$\\
$0.219$ & $57285.43$ & $-0.71827$ & $-0.73846$ & $-0.67429$\\
$0.219$ & $57285.60$ & $-0.71863$ & $-0.74078$ & $-0.67175$\\
$0.295$ & $57313.09$ & $0.70989$ & $0.27094$ & $-0.96260$\\
$0.394$ & $57349.13$ & $-0.67941$ & $-0.94664$ & $0.32228$\\
$0.454$ & $57371.31$ & $0.66163$ & $-0.70277$ & $-0.71142$\\
$0.454$ & $57371.38$ & $0.66299$ & $-0.70235$ & $-0.71183$\\
$0.555$ & $57408.01$ & $-0.67387$ & $-0.17397$ & $0.98475$\\
$0.555$ & $57408.09$ & $-0.67267$ & $-0.17437$ & $0.98468$\\
$0.618$ & $57431.01$ & $0.68219$ & $-0.96352$ & $0.26763$\\
$0.714$ & $57466.30$ & $-0.70874$ & $0.76098$ & $0.64877$\\
$0.794$ & $57495.47$ & $0.70801$ & $-0.21053$ & $0.97759$\\
$0.884$ & $57528.53$ & $-0.68770$ & $0.94986$ & $-0.31269$\\
\end{longtable}
}

\end{document}